\documentclass[a4paper,12pt]{article}
\usepackage[utf8]{inputenc}
\usepackage[textwidth=18.0cm,textheight=25cm]{geometry}
\usepackage{authblk}
\usepackage{graphics}
\usepackage{relsize}
\usepackage{amsfonts,amssymb,amsmath}
\usepackage{graphicx,color,epsf,subfigure}
\usepackage[T1]{fontenc}
\usepackage[utf8]{inputenc}

\DeclareMathOperator{\Tr}{Tr}
\newcommand{\x}{{\boldsymbol{x}}}
\newcommand{\kk}{{\boldsymbol{k}}}
\newcommand{\q}[1]{\lq\lq#1\rq\rq}
\newcommand{\eq}[1]{(\ref{eq:#1})}  
\newcommand{\fig}[1]{\textbf{\ref{fig:#1}}}  
\newcommand{\Sec}[1]{\textbf{\ref{sec:#1}}}  

\newcommand{\diff}{\mathop{}\!\mathrm{d}}
\newcommand{\ii}{\textrm{i}\,}
\newcommand{\mins}{{m^2_\text{in}}_{}}
\newcommand{\mouts}{{m^2_\text{out}}_{}}
\newcommand{\oin}{\omega_\text{in}}
\newcommand{\oout}{\omega_\text{out}}
\newcommand{\op}{\omega_+}
\newcommand{\om}{\omega_-}

\newcommand{\gb}{\gamma_{\mathsmaller{B}}}
\newcommand{\gf}{\gamma_{\mathsmaller{F}}}
\newcommand{\GGE}{\rho_\mathsmaller{\text{GGE}}}

\title{\bf Momentum-space entanglement \\ after smooth quenches}
\author[1,2]{Daniel W. F. Alves\footnote{Email: \texttt{dwfalves@gmail.com}}}
\author[3]{Giancarlo Camilo\footnote{Email: \texttt{gcamilo@iip.ufrn.br}}}
\affil[1]{\footnotesize São Paulo State University (UNESP), Institute for Theoretical Physics (IFT) \break R. Dr. Bento T. Ferraz 271, Bl. II, 01140-070, Sao Paulo, SP, Brazil}
\affil[2]{\footnotesize Center for Quantum Mathematics and Physics, Department of Physics, University of California \break Davis CA 95616 USA}
\affil[3]{\footnotesize International Institute of Physics, Universidade Federal do Rio Grande do Norte\break Campus Universit\'ario, Lagoa Nova, Natal-RN 59078-970, Brazil}

\date{\vspace{-5ex}}

\begin{document}

\maketitle

\begin{abstract}
We compute the total amount of entanglement produced between momentum modes at late times after a smooth mass quench in free bosonic and fermionic quantum field theories. The entanglement and R\'enyi entropies are obtained in closed form as a function of the parameters characterizing the quench protocol. For bosons, we show that the entanglement production is more significant for light modes and for fast quenches. In particular, infinitely slow or adiabatic quenches do not produce any entanglement. Depending on the quench profile, the decrease as a function of the quench rate $\delta t$ can be either monotonic or oscillating. In the fermionic case the situation is more subtle and there is a critical value for the quench amplitude above which this behavior is changed and the entropies become peaked at intermediate values of momentum and of the quench rate. We also show that the results agree with the predictions of a Generalized Gibbs Ensemble and obtain explicitly its parameters in terms of the quench data. 
\end{abstract}

\section{Introduction}
\label{sec:intro}
\indent

A quantum quench is one of the simplest protocols to put a quantum system away from equilibrium. The typical setup is to prepare a state at some initial time slice (e.g., the ground state of a Hamiltonian $H_0$) and suddenly let it evolve in time with a different Hamiltonian $H_1$ which acts on this state in a non-trivial way. The problem can be equivalently formulated as that of a time-dependent Hamiltonian $H(t)$ that changes from $H_0$ to $H_1$ as one of its parameters change in time. This has the advantage of allowing smooth transitions that happen within a finite time scale $\delta t$ rather than instantaneously. In any case, one is usually interested in the dynamics of various physical observables such as correlation functions and entanglement measures during the process. The study of quenches provides a window to explore a number of important questions concerning the non-equilibrium quantum dynamics, such as the mechanism underlying thermalization (or not) of isolated systems \cite{Polkovnikov:2010yn}. It has attracted increasingly more attention due to recent developments in cold atom physics that made possible to experimentally probe the real-time dynamics following a quantum quench \cite{2008RvMP...80..885B}.

A special class of quenches that admits many exact results is that of mass quenches in free quantum field theories. The simplest example is that of a free massless boson with action
\begin{equation}\label{eq:massquenchintro}
 I=\frac{1}{2}\int \diff^dx\,\left[(\partial\phi)^2-m(t)^2\phi^2\right]\,,
\end{equation}
where $m(t)^2$ is a smooth function that asymptotes to constant values $\mins$ and $\mouts$ at $t=-\infty$ and $t=+\infty$, respectively, changing in time during a scale of roughly $\delta t$. This is to be seen as a simple representative of the class of generic quenches of scaling dimension $\Delta$ operators $\mathcal{O}_\Delta(x)$ in a conformal field theory (CFT) (here the massless scalar CFT, with $\mathcal{O}_\Delta=\phi^2$ and $\Delta=\frac{d-2}{2}$). Smooth mass quenches of this kind have been recently studied in the literature since certain mass profiles $m(t)$ are amenable to analytical solution for any quench rate $\delta t$. The main reason behind that is the observation that the problem can be equivalently understood as the one of a standard (constant mass) quantum scalar field placed in an cosmological background -- more precisely the Friedmann-Lem\^aitre-Robertson-Walker (FLRW) spacetime. The latter has been studied back in the 70's by Bernard and Duncan \cite{BERNARD1977201,PhysRevD.17.964}, where mode solutions for some specific choices of mass profiles were obtained and its quantization was performed, so the results together with the intuition from quantum field theory in curved spacetime can be adapted to understand the mass quenches above. This approach has been used in \cite{Das:2014jna,Das:2014hqa} to study the behaviour of 1-point functions following the quench (see also \cite{Buchel:2013gba,Das:2016lla,Dymarsky:2017awt} for related work). 

Apart from local quantities such as one-point functions, there are non-local quantities whose behaviour during a quantum quench is also of interest given that they can probe the dynamics at different distance scales. These include the entanglement entropy (EE)
\begin{equation}\label{eq:EEdef}
 S=-\Tr(\rho\log\rho)\,,
\end{equation}
where $\rho$ is the reduced density matrix of a given subsystem, or the closely related one-parameter family of R\'enyi entropies
\begin{equation}\label{eq:Renyidef}
 S^{(q)}=\frac{1}{1-q}\log\Tr(\rho^q)
\end{equation}
with $0\le q<\infty$ ($q\ne1$), which includes the EE as the limiting case $q\to1$. 
Their time evolution has been thoroughly studied in the literature for the case of spatial subregions (i.e., where the system is split in such a way that the degrees of freedom live either inside or outside a given spatial region) in a variety of models using CFT techniques \cite{2005JSMTE..04..010C}, numerical simulations \cite{2017NatPh..13..246K}, and holography \cite {2011NJPh...13d5017A}, \cite{2014PhRvL.112a1601L}. The study of scaling properties of the EE as a function of the quench rate $\delta t$ was recently initiated in \cite{Caputa:2017ixa} for the harmonic chain, where the authors found consistence with the fast quench scaling behaviour of correlation functions and with Kibble-Zurek scaling in the appropriate regimes.    

In the present paper, we focus instead on \textit{momentum-space entanglement}. That is, we divide the Fock space in terms of positive and negative-momentum modes of the quantum field and calculate the entanglement production due to the quench between a single mode and all the others (though we shall see that only modes carrying opposite momenta are actually entangled). We investigate mass quenches in both free scalar and free fermion theories. The work follows the logic of \cite{Ball:2005xa} (see also \cite{Fuentes:2010dt}), which discussed the entanglement produced between momentum modes due to the expansion of the universe for a specific choice of FLRW scale factor. We will consider the same choice, which in our case translates to mass-increasing or mass-decreasing quenches, as well as another class of quench profiles that recovers the initial mass at late times. For previous work on momentum space entanglement in quantum field theory we refer the reader to \cite{Balasubramanian:2011wt,Grignani:2016igg,Hsu:2012gk}; see also \cite{PhysRevLett.113.256404,PhysRevB.94.081112} for similar work with spin chains.


An important aspect is that, 
unlike the majority of interacting quantum systems which are known to thermalize after a quench in the sense of approaching a thermal (or Gibbs) ensemble at late times, integrable systems such as the free field models studied here do not reach thermal equilibrium in this usual sense. This is because they possess an infinite number of conserved charges that conspire to constrain the dynamics, a fact that was first observed experimentally in \cite{2006NatureKinoshita} using a system of Bose gases in one dimension. However, later in \cite{2007PhRvL..98e0405R} it was proposed that integrable systems do thermalize in a broader sense to a new kind of equilibrium state called the Generalized Gibbs Ensemble (GGE) (see, e.g., \cite{2016JSMTE..06.4007V} for a review and validity checks for a variety of 1D systems). Having that in mind, we will also show that our results for the entanglement production precisely match the GGE prediction and, moreover, that the parameters characterizing this ensemble can be expressed in closed form in terms of the parameters defining the quench protocol. This reinforces the relevance of considering momentum-space entanglement as an interesting probe of thermalization of quantum systems. Unlike its real-space counterpart, the entanglement between momentum modes is not tied to any particular spatial sub-region of the system, so one should expect it to capture different physics characterized by non-local correlations in real space. It is worth to recall that the most accepted explanation for the spreading of real-space entanglement after a quench relies on the quasi-particle picture introduced in \cite{2005JSMTE..04..010C}, where it is assumed that pairs of quasi-particles with opposite momenta created within the correlation length are entangled, while those far apart from each other are not. As these particles travel through the system, real-space entanglement happens to grow ballistically. This intuition is able to correctly reproduce many of the qualitative features of the entanglement dynamics, and has been used to derive new results \cite{Alba:2017lvc}. From our calculations we can make this concrete by computing exactly how entangled the different particles produced are as a function of the parameters controlling the quench, such as its amplitude and speed. Hence, understanding the dynamics of momentum-space entanglement may also shed light into our understanding of how real-space entanglement grows. The free field example is chosen for convenience, since it is amenable to exact analytical results while still allowing for interesting phenomena such as the approach to the GGE, but we hope that our study can be a useful benchmark in future studies of thermalization in more complicated interacting models. 

The paper is organized as follows. In Section {\Sec{review}, we review the exact mode solutions and quantization of a free massive scalar field in FLRW spacetime and show how a conformal rescaling provides the solution to a mass quench in flat spacetime. In Section \Sec{EEscalar} we calculate the total amount of momentum space entanglement produced by different quenches at late times by following the same logic used in \cite{Ball:2005xa} for the curved space picture. Section \Sec{fermionic} presents the generalization for a fermionic field, while Section \Sec{GGE} compares the results with the prediction using a generalized Gibbs ensemble and shows an explicit expression for it in terms of quench parameters. Finally, Section \Sec{conclusions} contains the closing remarks.

\section{Smooth quantum quenches from FLRW fields}
\label{sec:review}
\indent

We begin by reviewing the connection between quantum quenches and fields in an expanding spacetime initially raised in \cite{Das:2014jna,Das:2014hqa}. A simple way to think of it is to start with the action for a real scalar field with mass $m$ conformally coupled to a curved background metric $g_{\mu\nu}(x)$ in any number $d$ of spacetime dimensions,
\begin{equation}\label{eq:scalarfieldaction}
 I=\frac{1}{2}\int \diff^dx\sqrt{-g}\ \big(g^{\mu\nu}\nabla_\mu\varphi\nabla_\nu\varphi-m^2\varphi^2-\xi R\varphi^2\big)\,.
\end{equation}
Here, $R=R[g]$ is the Ricci scalar curvature associated with $g_{\mu\nu}$. This gives rise to the equation of motion
\begin{equation}\label{eq:KGeq}
 \big(\Box_g+m^2+\xi R[g]\big)\varphi=0\,.
\end{equation}
For the special choice of parameter
\begin{equation}\label{eq:conformalcoupling}
 \xi=-\frac{d-2}{4(d-1)}\,
\end{equation}
it is well-known \cite{Birrell:1982ix} that the action in the massless case ($m=0$) is invariant under the Weyl rescaling  
\begin{eqnarray}\label{eq:Weyltransf}
 g_{\mu\nu}(x)&\to& \Omega(x)^2g_{\mu\nu}(x)\notag\\
 \varphi(x)   &\to& \Omega(x)^{-\Delta}\varphi(x) \qquad\text{with}\quad \Delta=\frac{d-2}{2}\,.
\end{eqnarray}
At the level of the equation of motion this property is nicely summarized by the following operator identity
\begin{equation}\label{eq:KGidentity}
 \Box_{g}+\xi R[g] \quad\longrightarrow\quad \Omega^{-\Delta-2}\left(\Box_{g}+\xi R[g]\right)\Omega^{\Delta}\,,
\end{equation}
which shows that when acting on a scalar field that transforms simultaneously as shown in \eq{Weyltransf} the equation is left invariant.

Such a symmetry is particularly useful for the purposes of quantization of the field when the curved background $g_{\mu\nu}(x)$ is a conformally flat spacetime, i.e., $g_{\mu\nu}(x)=\Omega(x)^2\eta_{\mu\nu}$, since one can use a Weyl rescaling to map to the Minkowski metric $\eta_{\mu\nu}$ and then resort to intuition from canonical field quantization in flat spacetime. 

The conformally flat background we shall be interested in is the FLRW spacetime with vanishing spatial curvature. In terms of the conformal time $t$ the metric can be written as $g_{\mu\nu}(t)=a(t)^2\eta_{\mu\nu}$, i.e,
\begin{equation}\label{eq:FRWmetric}
 ds^2 = a(t)^2\big[-\diff t^2+\diff\x^2\big]\,,
\end{equation}
which describes a spatially homogeneous and isotropic geometry that expands according to the scale factor $a(t)$. Unlike the massless case discussed above, the massive scalar field action \eq{scalarfieldaction} does not remain invariant under Weyl rescalings. In fact, by introducing the new scalar field $\phi$ according to $\varphi(t,\x)\equiv a(t)^{-\Delta}\phi(t,\x)$ the action can be rewritten in terms of the flat spacetime metric as 
 \begin{equation}\label{eq:quenchaction}
 I=\frac{1}{2}\int \diff^dx\ \big(\eta^{\mu\nu}\partial_\mu\phi\partial_\nu\phi-m(t)^2\phi^2\big)\,
\end{equation}
(there is no Ricci scalar contribution since $R[\eta]=0$) where 
\begin{equation}\label{eq:massquenchgeneral}
 m(t)^2\equiv m^2 a(t)^2\,.
\end{equation}
This is nothing but a quantum quench $m(t)$ in the mass of a scalar field in flat spacetime. Hence we see that the problem of mass quenches in flat space can be recast as the \q{dual} or equivalent problem of a (constant mass) conformally coupled quantum scalar field $\varphi$ living in a curved FLRW background.\footnote{Actually in $d=2$ the conformal coupling \eq{conformalcoupling} vanishes, so the FLRW scalar field has the usual minimal coupling.} 

Of course such a relation is not a peculiarity of mass quenches. It is a straightforward exercise to check that a general interaction term $\sim \lambda\varphi^n$ ($n\ge2$) in the action \eq{scalarfieldaction} in FLRW gets mapped under the same Weyl rescaling to a quench $\sim\lambda(t)\phi^n$ of the transformed scalar field that lives in flat space, with $\lambda(t)\equiv \lambda\,a(t)^{n+(2-n)d/2}$. We are dealing here with the special case of $n=2$ only for practical reasons, since the equation of motion is linear.

In flat spacetime, the canonical procedure for quantizing fields \cite{Birrell:1982ix} begins with first finding the positive-frequency normal modes $u_\kk(t,\x)\sim e^{\ii(\kk\cdot\x-\omega t)}$ that solve the Klein-Gordon equation in order to express the quantum field $\phi$ in the basis formed by $u_\kk$ and its complex conjugate. Such a meaningful classification into positive-frequency modes, however, is only possible because flat spacetime admits a timelike Killing vector field $K\equiv\partial_t$ whose corresponding conservation law guarantees well-defined \q{energy} eigenvalues at any time, i.e., $\ii\partial_tu_\kk=\,\omega\,u_\kk$ with $\omega\ge0$. In FLRW spacetime this timelike isometry $K$ is clearly not present in general since the metric \eq{FRWmetric} depends explicitly on time, so the very first step of the quantization procedure seems to fail once we go beyond the flat space scenario. This is a general feature of curved spacetimes that can be boiled down to the diffeomorphism invariance of general relativity, namely the inexistence of a preferred time coordinate to which a meaningful notion of energy can be associated. 

Fortunately, this problem can be worked around at least when the conformal factor $a(t)^2$ in \eq{FRWmetric} asymptotically approaches constant values as $t$ goes to $\pm\infty$. In this case a timelike Killing vector emerges asymptotically in the past and future infinity and one can still make sense of the quantization procedure (i.e., asymptotic positive-frequency modes, the vacuum state, particle excitations, and so on). This particular subset of FLRW spacetimes will be the one of interest in the following. Besides this technical reason, incidentally it turns out to be a useful toy model from the point of view of mass quenches, where well-defined initial and final equilibrium states before and after the quench are required. It is important to stress though that it has little relevance from the perspective of cosmology, where none of the relevant scale factors $a(t)$ happens to saturate to constant values.

We will be interested in two representative cases where the mass profile $m(t)=m\,a(t)$ allows for exact solutions, but before particularizing to specific choices of $m(t)$ let us first sketch the general strategy (see next Section for explicit expressions in the cases of interest). 

The Klein-Gordon equation arising from \eq{quenchaction} is $\big[-\partial_t^2+\partial_\x^2+m(t)^2\big]\phi=0$. On the grounds of translation invariance in the spatial directions one can seek for normal modes of the Fourier form $u_\kk(t,\x)=(2\pi)^{-(d-1)/2}e^{\ii\kk\cdot\x}\chi_\kk(t)$, where $\kk\in\mathbb{R}^{d-1}$ is the spatial momentum.\footnote{Sometimes for a better handling of UV divergences it is convenient to restrict $\kk$ to a finite range such as a torus $\mathbb{S}^{d-1}$ of length $L$ (i.e., taking $k_i\in[0,L]$ with periodic boundary conditions at the two extrema). We will be implicitly doing this in the following when writing for instance Kronecker deltas $\delta_{\kk\kk'}$ instead of Dirac ones, which serves well our physical purposes here without unnecessary technical complications of dealing with infinities.} The mode functions $\chi_\kk(t)$ are easily shown to satisfy a simple harmonic oscillator equation with a time-dependent fundamental frequency, namely
\begin{equation}\label{eq:modeeqoscillator}
\frac{d^{2}\chi_\kk}{dt^2}+\big[\kk^2+m(t)^2\big]\chi_\kk = 0\,.
\end{equation}
Different solutions can be found depending on the boundary conditions imposed. The ones behaving as positive-frequency $\sim \frac{1}{\sqrt{2\oin}}e^{-\ii\oin t}$ at early times $t\to-\infty$ are referred to as \q{pre-quench} modes or simply \q{in} modes 
$\chi^{\text{in}}_\kk(t)$ (borrowing terminology from the curved space description), while the ones that behave as positive-frequency $\sim \frac{1}{\sqrt{2\oout}}e^{-\ii\oout t}$ at late times $t\to+\infty$ are called \q{post-quench} or \q{out} modes $ \chi^{\text{out}}_\kk(t)$. 
Both constitute equally good basis functions in terms of which the field $\phi$ can be expressed, i.e.,
\begin{align}\label{eq:bogofield}
 \phi(t,\x) &= \int \frac{d^{d-1}\kk}{(2\pi)^{d-1}}\,e^{\ii\kk\cdot\x}\big[a^\text{in}_\kk\chi^{\text{in}}_\kk(t)+a^{\dagger\,\text{in}}_{-\kk}\chi^{*\,\text{in}}_{-\kk}(t)\big] 
 =  \int \frac{d^{d-1}\kk}{(2\pi)^{d-1}}\,e^{\ii\kk\cdot\x}\big[a^\text{out}_\kk\chi^{\text{out}}_\kk(t)+a^{\dagger\,\text{out}}_{-\kk}\chi^{*\,\text{out}}_{-\kk}(t)\big]\,,
\end{align}
so they must be related by a Bogoliubov transformation. Translational invariance of the model in the spatial directions ensures that this transformation is of the following block-diagonal form 
\begin{eqnarray}\label{eq:bogomodes}
 \chi^{\text{in}}_\kk &=& \alpha_\kk\ \chi^{\text{out}}_\kk + \beta_\kk\ \chi^{*\,\text{out}}_{-\kk}
\end{eqnarray}
with coefficients $\alpha_\kk\,,\beta_\kk$ to be determined. Accordingly, it follows from \eq{bogofield} and \eq{bogomodes} that the operators $a^{}_\kk\,,a^\dagger_\kk$ in both bases relate as
\begin{eqnarray}\label{eq:bogoa}
 a^{\text{in}}_\kk &=& \alpha^*_{-\kk}\ a^{\text{out}}_{\kk} - \beta^*_{-\kk}\ a^{\dagger\,\text{out}}_{-\kk}\notag\\
 a^{\text{out}}_\kk &=& \alpha_{\kk}\ a^{\text{in}}_{\kk} + \beta^*_{-\kk}\ a^{\dagger\,\text{in}}_{-\kk}\,.
\end{eqnarray}

The quantization of $\phi$ then proceeds by postulating the standard creation/annihilation algebra for the operators $a^{}_\kk\,,a^\dagger_\kk$ (either \q{in} or \q{out} ones),
\begin{equation}
 \big[a^{}_\kk,a^\dagger_{\kk'}\big]=\delta_{\kk\kk'}\,,\qquad\big[a_\kk,a_{\kk'}\big]=\big[a^\dagger_\kk,a^\dagger_{\kk'}\big]=0\,.
\end{equation}
Annihilation by the operator $a^{\text{in}}_\kk$ defines the vacuum state $|0_\text{in}\rangle$ interpreted as the state with no particles as seen by an observer at $t=-\infty$, or before the quench, i.e.,
\begin{equation}
 a^{\text{in}}_\kk|0_\text{in}\rangle = 0\qquad\forall\kk\,.
\end{equation}
Acting on it with $a^{\dagger\,\text{in}}_\kk$ we can construct particle states of the \q{in} type carrying momenta $\kk$ and energy $\oin$. Similarly, annihilation by $a^{\text{out}}_\kk$ defines another vacuum state $|0_\text{out}\rangle$ that now has no particles according to an observer at $t=+\infty$ or after the quench,
\begin{equation}
 a^{\text{out}}_\kk|0_\text{out}\rangle = 0\qquad\forall\kk\,,
\end{equation}
from which we obtain particle states of the \q{out} type carrying momenta $\kk$ and energy $\oout$ through the action of $a^{\dagger\,\text{out}}_\kk$. 

The crucial fact here with no analog in the constant mass case is that even the notion of particles in the present case is time-dependent, not universal. From the equivalent point of view of quantum fields in FLRW spacetime this is just the well-accepted statement that the definition of particles is observer-dependent in a curved space. 
Intuitively, this property is already manifested above in the fact that particle excitations experienced by the two different asymptotic observers carry different energies $\oin\ne\oout$, or by the fact that $|0_\text{out}\rangle$ is \emph{not} annihilated by $a^{\text{in}}_\kk$ (and vice-versa). Indeed, a no-particle state for one observer can look like a complicated particle bath as told by the other. For instance, the number of particle excitations carrying momentum $\kk$ counted by an asymptotic observer at past infinity (using the number operator $N^\text{in}_\kk\equiv a^{\dagger\,\text{in}}_\kk a^{\text{in}}_\kk$) obviously vanishes in the \q{in} vacuum, $\langle0_\text{in}|N^{\text{in}}_\kk|0_\text{in}\rangle=0$, while (by virtue of \eq{bogoa}) it is non-zero in the \q{out} vacuum, 
\begin{equation}\label{eq:particleproduction}
 \langle0_\text{out}|N^{\text{in}}_\kk|0_\text{out}\rangle=|\beta_\kk|^2\,.
\end{equation}
This phenomenon is often referred in the general relativity literature as particle production by the gravitational field. In the quench picture it states that, from the point of view of the initial observer, the full process of quenching the scalar field mass from $m_\text{in}$ to $m_\text{out}$ has produced a bath of $|\beta_\kk|^2$ particles carrying momentum $\kk$.

It is interesting to note that after using the mode expansion \eq{bogofield} the momentum space Hamiltonian describing the mass quench above reads
\begin{eqnarray}\label{eq:Hk}
 H = \frac{1}{2}\int d^{d-1}\kk\,\left[2\big(|\dot\chi_\kk|^2+\omega_\kk^2|\chi_\kk|^2\big)a^\dagger_{\kk}a_\kk + \big(\dot\chi_\kk^2+\omega_\kk^2\chi_\kk^2\big)a_\kk a_{-\kk} + \big({\dot\chi}^*_\kk\,\!^2+\omega_\kk^2\chi^*_\kk\,\!^2\big)a^\dagger_\kk a^\dagger_{-\kk} \right]+E_0\,,
\end{eqnarray}
where $\chi_\kk$ here can be either the \q{in} or \q{out} modes, $E_0\equiv\frac{1}{2}\int d^{d-1}\kk\,\big(|\dot\chi_\kk|^2+\omega_\kk^2|\chi_\kk|^2\big)$ is the vacuum energy contribution, and $\omega_\kk(t)^2\equiv \kk^2+m(t)^2$. We see that, in the presence of the quench $m(t)$, in addition to the usual particle number piece $\sim a^\dagger_\kk a_\kk$ the Hamiltonian contains also interaction terms between opposite momentum modes $\kk$ and $-\kk$. Such terms disappear when $m(t)=\text{const}$, where the modes are simply $\chi_\kk(t)=\frac{1}{\sqrt{2\omega_\kk}}e^{-\ii\omega_\kk t}$ and $H$ reduces to its standard form $H=\int d^{d-1}\kk\,\omega_\kk a^\dagger_{\kk}a_\kk+E_0$. 

Equation \eq{Hk} suggests that it might be interesting to split the Fock space into $\kk$ and $-\kk$ modes and calculate the entanglement properties of this bipartite system. This shall be done in Section \Sec{EEscalar}, but first let us illustrate the results above explicitly for the two cases of interest.

\subsection{Two specific quench profiles}
\subsubsection{Tanh profile}
\indent\par

The first quench profile of interest is
\begin{equation}\label{eq:massquenchtanh}
 m(t)^2 = \frac{1}{2}\left(\mouts+\mins\right)+\frac{1}{2}\left(\mouts-\mins\right)\tanh \frac{t}{\delta t}\,,
\end{equation}
which smoothly interpolates (during a time scale of roughly $\delta t$) from an initial \q{pre-quench} value $\mins$ at $t=-\infty$ to another final \q{post-quench} value $\mouts$ at $t=+\infty$. This is reminiscent of the massive scalar field model studied in \cite{BERNARD1977201,PhysRevD.17.964}, defined on a FLRW background with scale factor $a(t)^2=A+B\tanh \tfrac{t}{\delta t}$, where the cosmological parameters $A$ and $B$ are related to the quench data by
\begin{equation}\label{eq:minmoutdef}
 \mins = m^2(A-B)\qquad \mouts = m^2(A+B)\,.
\end{equation}
The extreme limit of $\delta t\to\infty$ corresponds to an infinitely slow or \emph{adiabatic} quench, while $\delta t\to0$ would correspond to a step function or \emph{instantaneous} quench profile of the type discussed in \cite{2007JSMTE..06....8C,Das:2015jka}. For $\mouts=\mins$ (or $B=0$) we recover the static case of no quench.

The in and out mode solutions for this system were originally obtained in \cite{Bernard_Duncan_1977} and are given by
\begin{align}
 \chi^{\text{in}}_\kk(t) &= \frac{1}{\sqrt{2\oin}}e^{-\ii\op t-\ii\om\delta t \log(2\cosh\frac{t}{\delta t})}\, _2F_1\left[1+\ii\om \delta t;\,\ii\om \delta t;\,1-\ii\oin \delta t;\,\frac{1+\tanh\frac{t}{\delta t}}{2}\right]\notag\\
 \chi^{\text{out}}_\kk(t) &= \frac{1}{\sqrt{2\oout}}e^{-\ii\op t-\ii\om\delta t \log(2\cosh\frac{t}{\delta t})}\, _2F_1\left[1+\ii\om \delta t;\,\ii\om \delta t;\,1+\ii\oout \delta t;\,\frac{1-\tanh\frac{t}{\delta t}}{2}\right]\,,
\end{align}
where $_2F_1[\cdots]$ is the hypergeometric function and we have introduced the shorthand notation
\begin{eqnarray}\label{eq:omegasdef}
 \omega_\text{in}=\sqrt{\kk^2+\mins}\,,\qquad \omega_\text{out}=\sqrt{\kk^2+\mouts}\,,\qquad\text{and}\qquad \omega_\pm=\frac{\oout\pm\oin}{2}\,.
\end{eqnarray}
By working out well-known algebraic properties of the hypergeometric functions that convert the argument $z$ into $1-z$ (here $z=\frac{1}{2}(1+\tanh\frac{t}{\delta t})$), the Bogoliubov coefficients \eq{bogomodes} for the present model have been obtained in \cite{PhysRevD.17.964}, namely
\begin{eqnarray}\label{eq:bogocoeffs}
 \alpha_\kk &=& \sqrt{\frac{\oout}{\oin}}\,\frac{\Gamma(1-\ii\oin\delta t)\,\Gamma(-\ii\oout\delta t)}{\Gamma(-\ii\op\delta t)\,\Gamma(1-\ii\op\delta t)}\notag\\
 \beta_\kk &=& \sqrt{\frac{\oout}{\oin}}\,\frac{\Gamma(1-\ii\oin\delta t)\,\Gamma(\ii\oout\delta t)}{\Gamma(\ii\om\delta t)\,\Gamma(1+\ii\om\delta t)}\,.
\end{eqnarray}
In particular, their absolute values are easily checked to take the simple forms
\begin{eqnarray}\label{eq:bogocoeffssquared}
 |\alpha_\kk|^2 &=& 1+|\beta_\kk|^2 = \frac{\sinh^2(\pi\op\delta t)}{\sinh(\pi\oin\delta t)\sinh(\pi\oout\delta t)}\,.
\end{eqnarray}

An interesting limiting case is that of abrupt quenches ($\delta t\to0$), where the Tanh profile \eq{massquenchtanh} becomes the step function $m(t)^2 = \mins\theta(-t)+\mouts\theta(t)$ and the in and out modes are simple plane waves with frequency $\oin$ and $\oout$, respectively. In this case the coefficients above simplify to
\begin{align}\label{eq:bogocoeffinstant}
\alpha_\kk^\text{instant} = \frac{\omega_+}{\sqrt{\oin\oout}}\,, \quad \beta_\kk^\text{instant} = \frac{\omega_-}{\sqrt{\oin\oout}}\,.
\end{align}

\subsubsection{Sech profile}
\indent\par

The second quench of interest is the Gaussian-like profile 
\begin{align}\label{eq:massquenchsech}
m(t)^2 = m_{0}^{2}\,\text{sech}^{2}\Big(\frac{t}{\delta t}\Big)\,,
\end{align}
which has the same initial and final values of mass $m_0^2$ and was studied previously in \cite{Das:2014hqa}. The in and out solution read \cite{nova_ref_sech}
\begin{align}
\chi_{\kk}^{\text{in/out}}(t) &= e^{-\ii\!kt}\left(1+e^{2t/\delta t}\right)^{\mu}\left[c_{1}^{\text{in/out}}e^{2\ii\!kt}{}_{2}F_{1}\left(\mu,\ii\!k\delta t+\mu,1+\ii\!k\delta t,-e^{2t/\delta t}\right)\notag\right.\\
&\left.+c_{2}^{\text{in/out}}{}_{2}F_{1}\left(\mu,-\ii\!k\delta t+\mu,1-\ii\!k\delta t,-e^{2t/\delta t}\right)\right]\,,
\end{align}
where we have defined
\begin{equation}\label{eq:mu}
\mu=\frac{1+\sqrt{4m_{0}^{2}\delta t^2+1}}{2}\,.
\end{equation}

The Bogoliubov coefficients relating in and out solutions are given by \cite{nova_ref_sech}
\begin{align} \label{eq:bogocoeffssquaredsech}
\alpha\left(k\right) &= \frac{\Gamma(1+\ii\!k\delta t)\Gamma(\ii\!k\delta t)}{\Gamma(\ii\!k\delta t-\mu+1)\Gamma(\ii\!k\delta t+\mu)}\\
\beta(k) &= \ii\sin\left(\pi\mu\right)\text{csch}\left(\pi k\delta t\right)
\end{align}
 We will use these expressions to compute the entanglement and Rényi entropies below.

\section{Entanglement production by the quench}
\label{sec:EEscalar}
\indent

The goal of the present section is to study the total amount of entanglement produced between scalar field modes by the mass quenching process introduced above. The discussion follows closely the one of \cite{Ball:2005xa}. 
In what follows, we will be working in the Heisenberg picture, so states are not supposed to change in time. 

Let us first give a qualitative picture of what is going on here before delving into the calculations. We have seen in last section that there are two equally good bases in which one can express the field, one adapted to early and the other to late time observers. In the cosmological problem, both observers use the same time coordinate $t$  (defined by the metric \eq{FRWmetric}) since its tangent vector provides a timelike Killing vector adapted to each of them. Given this time coordinate, both observers can then define a local notion of particle excitation and vacuum, which is valid only at either the asymptotic past or future. That is why, even by working in the Heisenberg picture, we are able to talk about particle and entanglement production after the expansion of the universe. 

The same reasoning translates directly to the problem of quantum quenches, which does not involve curved spacetimes at all (it is defined in flat spacetime) but we know to be equivalent to a FLRW field. Now, both the pre-quench and post-quench observers use the same time coordinate $t$ associated to some inertial reference system. But if an experimenter preparing the state at early times is to have a meaningful notion of particles, he or she must use the in-modes, while an observer who will analyze the system at late times, after the quench is finished, naturally picks the out-modes. As a result, we will now show that an initial product state (with respect to positive and negative-momentum bipartitioning of the Fock space) is seen by the post-quench observer as entangled. This entanglement production is obviously tied to our time-dependent Hamiltonian (even though the states do not evolve in time), since the in and out solutions are obtained from it. 

We now provide the details for the argument above. As suggested by \eq{Hk}, opposite momentum modes $\kk$ and $-\kk$ provide a natural splitting of the Fock space which can be used to explore entanglement properties of the model. For simplicity it is assumed that the initial state is the vacuum $|{0_\text{in}}\rangle$ of the Hamiltonian with mass $\mins$, which can be thought of as the state\footnote{Strictly speaking, the product over momentum modes above is not well-defined since the momentum $\kk$ is a continuum variable. The way to make sense of it is by restricting $\kk$ to a compact space (e.g., using periodic boundary conditions) as mentioned previously. Anyway, none of the conclusions below are affected by this subtlety.}
\begin{equation}\label{eq:vacuuminbipartite}
 |{0_\text{in}}\rangle
 =\bigotimes_{\kk}|0_{\text{in}_\kk} 0_{\text{in}_{-\kk}}\rangle
\end{equation}
having zero excitation number in any of the momentum modes (as told by the pre-quench observer).\footnote{Here we are using the shorthand notation $|n_{\text{in}_\kk}\,n_{\text{in}_{-\kk}}\rangle\equiv|n_{\text{in}_\kk}\rangle\otimes|n_{\text{in}_{-\kk}}\rangle$ for tensor products between states belonging to subspaces with opposite momenta $\pm \kk$ (and similarly for \q{out} states).} 

Since the Bogoliubov transformation \eq{bogomodes} only mixes opposite momenta $\kk$ and $-\kk$, each fixed $\kk$ piece of \eq{vacuuminbipartite} admits the following Schmidt decomposition in terms of the out basis
\begin{equation}\label{eq:vacuumintoout}
|0_{\text{in}_\kk}\,0_{\text{in}_{-\kk}}\rangle=\sum^\infty_{n=0}c_n|n_{\text{out}_\kk}\,n_{\text{out}_{-\kk}}\rangle\,,
\end{equation}
where the coefficients are all real and $n_\text{out}$ labels the excitation number according to a post-quench observer. 
An explicit expression for the $c_n$ can be obtained by applying \eq{bogoa} to \eq{vacuumintoout}, i.e., (notice from \eq{bogocoeffs} that $\alpha_{-\kk}=\alpha_\kk$ and $\beta_{-\kk}=\beta_\kk$)
\begin{equation}
0=a^{\text{in}}_\kk|0_{\text{in}_\kk}\,0_{\text{in}_{-\kk}}\rangle=
\sum^\infty_{n=0}(\alpha^*_{\kk}c_{n+1}-\beta^*_{\kk}c_n)\sqrt{n_\text{out}+1}\,|n_{\text{out}_\kk}\,(n_\text{out}+1)_{-\kk}\rangle\,,
\end{equation}
which is solved by
\begin{equation}\label{eq:Schmidtcn}
c_n=\left(\frac{\beta^*_{\kk}}{\alpha^*_{\kk}}\right)^n c_0\,.
\end{equation}
The remaining coefficient $c_0$ is fixed by requiring \eq{vacuumintoout} to have unit norm, namely
\begin{equation}\label{eq:Schmidtc0}
c_0=\sqrt{1-\left|\frac{\beta_{\kk}}{\alpha_{\kk}}\right|^2}.
\end{equation}
Therefore we see that the in-vacuum \eq{vacuuminbipartite}, which was perceived at initial times as a simple product (i.e., unentangled) state of opposite-momentum modes, is seen by a post-quench observer as a highly entangled state made of infinitely many particle excitations of the out type, with the Schmidt coefficients determined by the Bogoliubov ones $\alpha_\kk,\beta_\kk$. 

We can now proceed to quantify how much entanglement has been produced by the quench. We focus on a single pair of modes ($\kk,-\kk$), since opposite-momenta are the only case allowed by the Bogoliubov mode mixing and the result for multiple pairs of modes is easily obtained from this. 
The density matrix for our bipartite system of opposite momentum modes is therefore\footnote{The proper way to construct it is by tracing out all momentum modes in the initial vacuum \eq{vacuuminbipartite} except $\kk$, which is a trivial operation giving an overall factor of unity for each mode and yielding \eq{rhototal} at the end.
}
\begin{equation}\label{eq:rhototal}
 \rho = |0_{\text{in}_\kk}\,0_{\text{in}_{-\kk}}\rangle\langle0_{\text{in}_\kk}\,0_{\text{in}_{-\kk}}| = \sum^\infty_{n=0}|c_n|^2\,|n_{\text{out}_\kk}\,n_{\text{out}_{-\kk}}\rangle\langle n_{\text{out}_\kk}\,n_{\text{out}_{-\kk}}| \,.
\end{equation}
This will be the state at any time and the physical properties of pre- or post-quench stages are entirely encoded in which basis of the Hilbert space we decide to use. 
In order to obtain the entanglement produced between the modes at the end of the process, we need to trace out \emph{with respect to the out-basis} the negative-momentum modes to obtain the reduced state
\begin{eqnarray}\label{eq:rhok}
 \rho_\kk &\equiv& \Tr_{\{-\kk\}}\rho 
 = \sum^\infty_{n=0}|c_n|^2\,|n_{\text{out}_\kk}\rangle\langle n_{\text{out}_\kk}|\,
\end{eqnarray}
describing the positive-momentum ones. Notice that had we traced out with respect to the in-basis we would have obtained the unentangled state $\rho_\kk=|0_{\text{in}_\kk}\rangle\langle 0_{\text{in}_\kk}|$.

The R\'enyi entropies \eq{Renyidef} can be used to quantify the amount of entanglement between the modes $\kk$ and $-\kk$. They are constructed from the reduced state $\rho_\kk$ as
\begin{eqnarray}
 S^{(q)}_\kk = \frac{1}{1-q}\log\Tr_{\{\kk\}}\rho_\kk^q = \frac{1}{1-q}\log\sum^\infty_{n=0}|c_n|^{2q}\,.
\end{eqnarray}
From \eq{Schmidtcn} and \eq{Schmidtc0} it is seen that $|c_n|^2$ depends on the index $n$ essentially as a power law, i.e.,
\begin{equation}
 |c_n|^2 = \gb^n(1-\gb)
\end{equation}
where we have introduced the parameter\,\footnote{The subscript $B$ stands for \q{boson}, to be contrasted with the fermionic case in the next Section.}
\begin{equation}\label{eq:gamma}
 \gb\equiv\left|\frac{\beta_{\kk}}{\alpha_{\kk}}\right|^2 \qquad (0\leq\gb<1) \,
\end{equation}
and used \eq{bogocoeffssquared} in the second equality to put it into a compact form. The summation over $n$ thus becomes just a geometric series which is easily carried out to yield the following closed-form expression for the Renyi entropies
\begin{equation}\label{eq:Renyi}
 S^{(q)}_\kk = \frac{1}{1-q}\log\frac{(1-\gb)^q}{1-{\gb}^q}\,.
\end{equation}
In particular, the limit $q\to1$ gives the entanglement entropy $S_\kk = -\Tr_{\{\kk\}}\big(\rho_\kk\log\rho_\kk\big)$, namely
\begin{equation}
\label{eq:EE}
S_\kk \equiv\lim_{q\to1}S^{(q)}_\kk= \log\frac{\gb^{\gb/(\gb-1)}}{1-\gb}\,.
\end{equation}
This agrees exactly with the expression originally obtained in \cite{Ball:2005xa} in the context of a scalar field in a cosmological setup. 

It is important to notice that tracing out the positive momentum modes instead of negative ones in \eq{rhok} would yield the same Renyi and entanglement entropies as long as the full density matrix $\rho$ is a pure state, since in this case the Schmidt form \eq{rhototal} guarantees the two reduced density matrices $\rho_{\kk}$ and $\rho_{-\kk}$ to have the same eigenvalues $|c_n|^2$.

The entropies \eq{Renyi} and \eq{EE} quantify the total amount of entanglement produced at late times by the quench. They depend on the quench rate $\delta t$, on the mode $\kk$, and on the masses $\mins,\mouts$ 
or $m_{0}^2$ for the Tanh or Sech profile, respectively. But the dependence on these four physical parameters is of a very peculiar kind: only combined inside the single parameter $\gb$ defined in \eq{gamma}. In this sense, $\gb$ contains all the information concerning the late-time entanglement between opposite-momentum modes. Its explicit form for the Tanh and Sech profiles introduced in Section \Sec{review} is readily found from \eq{bogocoeffssquared} and \eq{bogocoeffssquaredsech}, namely
\begin{align}\label{eq:gammas}
 \gb^{(\text{tanh})} &= \frac{\sinh^2(\pi\om\delta t)}{\sinh^2(\pi\op\delta t)}\notag\\
 \gb^{(\text{sech})} &= 
 \frac{2\sin^{2}\left(\pi\mu\right)}{\cosh\left(2k\pi\delta t\right)-\cos\left(2\pi\mu\right)}\,.
\end{align}

with $\mu$ given by \eq{mu}. Both $S_\kk^{(q)}$ and $S_\kk$ are monotonically increasing functions of $\gb$, so they can be inverted to yield $\gb(S)$, as noticed in \cite{Ball:2005xa}. This is rather interesting, meaning that the quench protocol $m(t)$ 
can in principle be fully reconstructed only from entanglement data. The expression for the EE is too complicated to invert and find $\gb(S)$ analytically, but for the Renyi's the situation is simple enough so that this can be done, namely, all we have to do is solve for $\gb$ the $q$-th order equation
\begin{equation}
(1-\gb)^q - s_q(1-\gb^q)=0\,,
\end{equation}
with $s_q\equiv e^{(1-q)S_\kk^{(q)}}$. The second Renyi entropy provides the simplest example, 
\begin{equation}
 \gb=\frac{1-e^{-S_\kk^{(2)}}}{1+e^{-S_\kk^{(2)}}}\,.
\end{equation}
In any case, having obtained $\gb=\gb(S)$ should be enough to express the quench parameters in terms of it using the definition \eq{gamma}. 

The physical interest, however, is on the $\kk$- and $\delta t$-dependence of the entropies themselves, which we now analyze in detail. 

\subsection{Tanh profile}
We start with the Tanh profile. Since $\gb$ is symmetric under $\oin\leftrightarrow\oout$ one can choose to focus on quenches that increase the mass, i.e., $\mouts>\mins$.\footnote{When $\mins=\mouts$, $\gb$ vanishes and there is no entropy production, which is trivial since in this case there is no quenching at all (see \eq{massquenchtanh})} An interesting special case is that of $\mins=0$, where the pre-quench Hamiltonian is that of a conformal field theory (the free massless boson). In the following we shall limit our numerical analysis to the entanglement entropy and the first few integer R\'enyi entropies ($q=2,3,4,5$).

\begin{figure}[htb]
\begin{center}
\includegraphics[width=.42\textwidth]{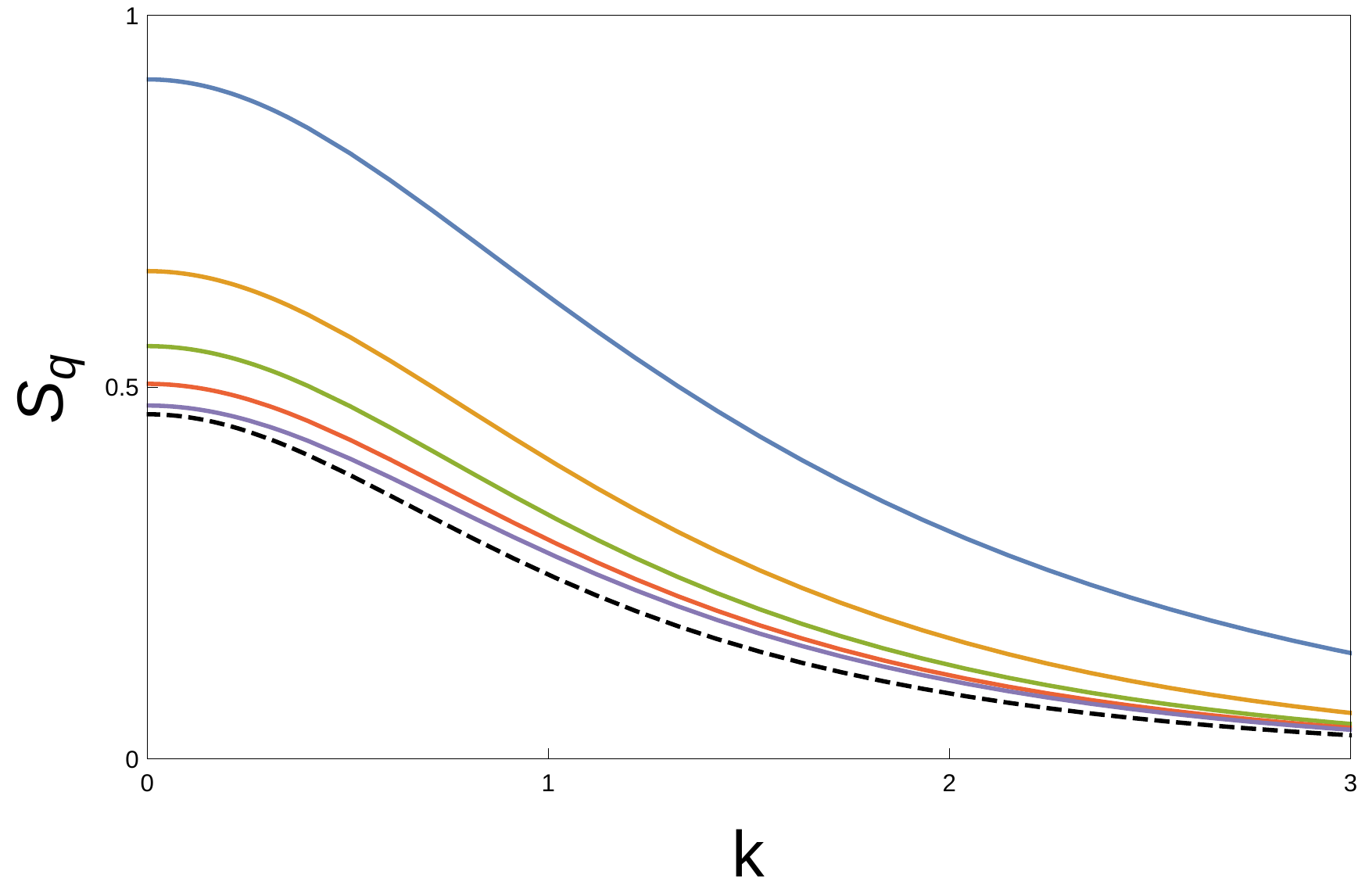}
\end{center}
\caption{{\small $k$-dependence of the entanglement entropy $S_\kk$ (blue top) and the R\'enyi entropies $S^{(q)}_\kk$ with $q=2,3,4,5$ (from top yellow to bottom purple) produced by the tanh mass quench for a fixed quench rate $\delta t=0.1$. The black dashed curve shows the corresponding particle production rate $|\beta_k|^2$ discussed in \eq{particleproduction}. In the plot we have set $\mins=1$ and $\mouts=4$, but the shape of the plots remains unchanged for other values. The magnitude of the entropies is controlled by the mass difference $\delta m^2=\mouts-\mins$, namely it grows as $\delta m^2$ is increased (the precise proportionality law is not known though).}}
\label{fig:EERenyifixeddt}
\end{figure}

For a fixed quench rate $\delta t$, all the entropies decrease monotonically with $k\equiv|\kk|$ as shown in Figure \fig{EERenyifixeddt}. This shows that more entanglement is produced between IR (low $k$) modes than between UV (high $k$) modes. The magnitude of the entropies is proportional to the mass difference $\delta m^2=\mouts-\mins$ between the initial and the final states. The maximal value corresponds to quenches that start from the massless boson CFT, although this case is subtle since there is formally a divergent zero mode contribution $\sim\log 1/k$ at $k=0$. In practice we can simply ignore this fact since in this case there is no ($\kk,-\kk$) splitting of modes at all to begin with. Let us recall that the usual upper bound $S_\mathsmaller{EE}\le\log\text{dim}(\mathcal{H})$ for the EE is infinite here since the Hilbert space for the reduced state $\rho_\kk$ is infinite-dimensional, so there is nothing contradictory about the fact illustrated in the plot that the EE is not limited from above.

\begin{figure}[htb]
\begin{center}
\includegraphics[width=.42\textwidth]{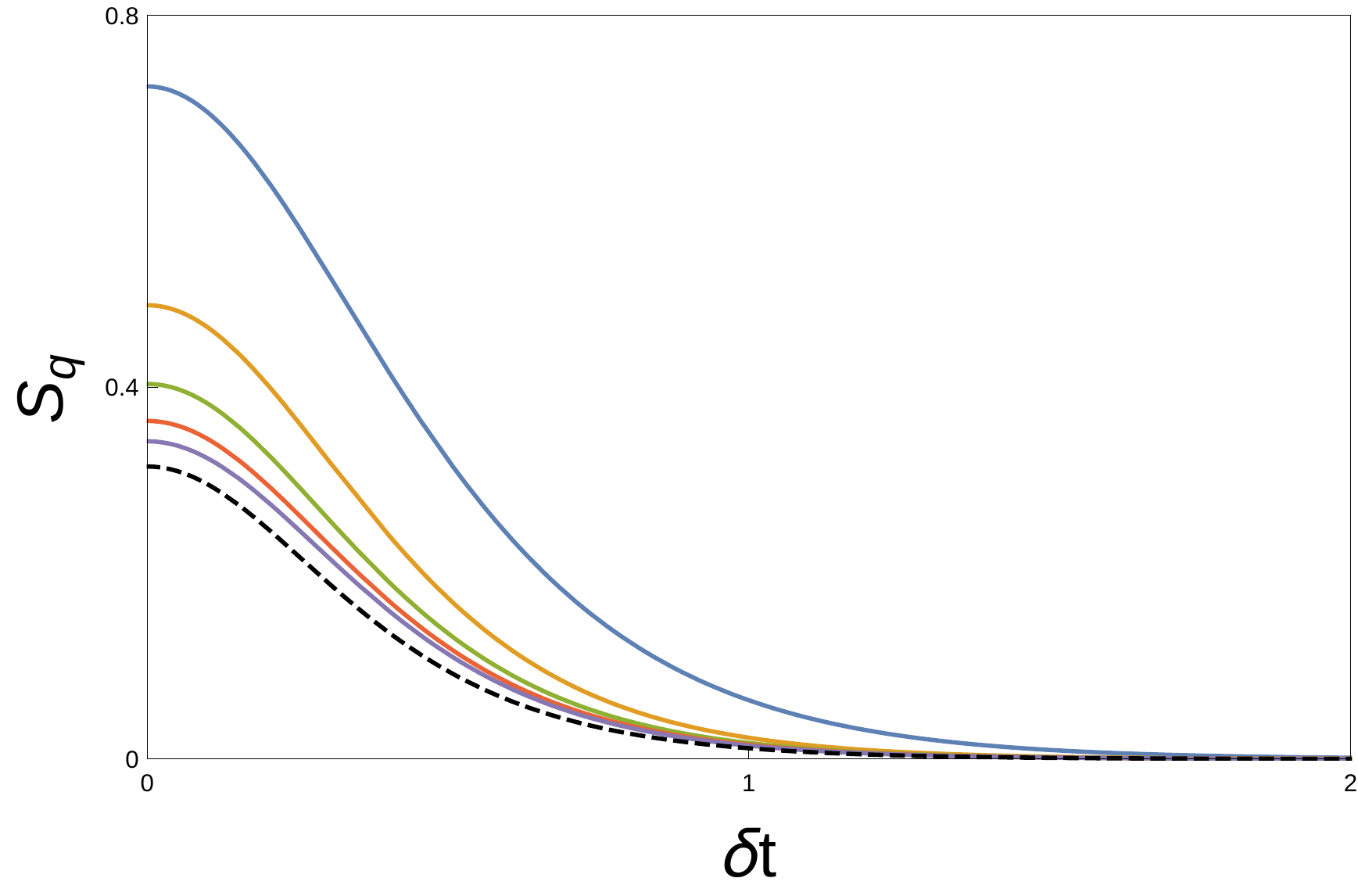}
\end{center}
\caption{{\small Entanglement entropy $S_\kk$ (blue) and the R\'enyi entropies $S^{(q)}_\kk$ with $q=2,3,4,5$ (from top yellow to bottom purple) as functions of the quench rate $\delta t$. The black dashed curve shows the corresponding particle production rate $|\beta_k|^2$ discussed in \eq{particleproduction}. For the plot we fix a single mode $k=0.5$ and $\mins=0.5,\mouts=2$, but the behavior is qualitatively the same for other values.}}
\label{fig:EERenyivarydt}
\end{figure}

Figure \fig{EERenyivarydt} illustrates the dependence of the entropies on the time scale $\delta t$ for a single mode $k$. It becomes clear that faster quenches produce more mode entanglement than slower ones. In particular, as $\delta t$ grows all curves approach zero asymptotically, indicating that infinitely slow or \emph{adiabatic quenches} (those with $\delta t\to\infty$) do not create any entanglement between field modes. 

\subsection{Sech profile}
For the case of the Sech profile, we plot the results on Figure \fig{EERenyifixeddtsech} and \fig{EERenyivarydtsech}. By fixing $\delta t$ and $m_{0}$, we see that all R\'enyi and entanglement entropies decrease monotonically with $k$. That is, light degrees of freedom are more entangled than heavy ones, similarly to the Tanh profile case. However,there is an interesting difference when fixing $k$ and exploring the $\delta t$ dependence, as shown in Figure \fig{EERenyivarydtsech}. The entanglement does not decrease monotonically as in the Tanh profile case, but rather oscillates with an amplitude that decreases as $\delta t$ increases. The origin of this oscillatory behavior can be seen in \eq{gammas}. Also, notice that again there is a divergence as $k$ approaches zero that we should not worry about as we commented above for the Tanh profile.

\begin{figure}[htb]
\begin{center}
\includegraphics[width=.42\textwidth]{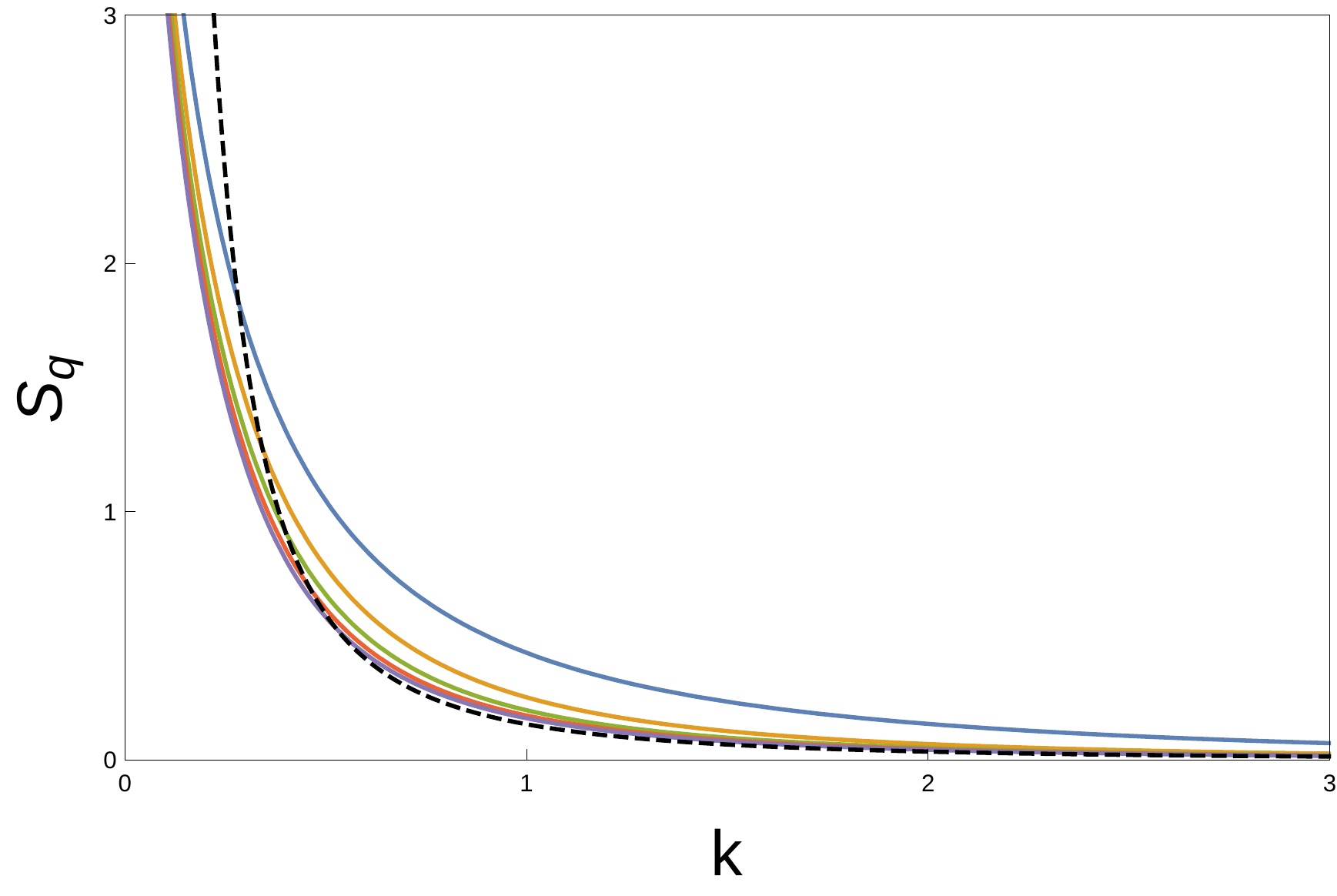}
\end{center}
\caption{{\small $k$-dependence of the entanglement entropy $S_\kk$ (blue) and the R\'enyi entropies $S^{(q)}_\kk$ with $q=2,3,4,5$ (from top yellow to bottom purple) produced by the sech mass quench for a fixed quench rate $\delta t=0.1$.  The black dashed curve shows the corresponding particle production rate $|\beta_k|^2$ discussed in \eq{particleproduction}. In the plot we have set $m_0=2$, but the shape of the plots remains unchanged for other values.}}
\label{fig:EERenyifixeddtsech}
\end{figure}

\begin{figure}[htb]
\centering
    \subfigure[]{\includegraphics[width=.4\textwidth]{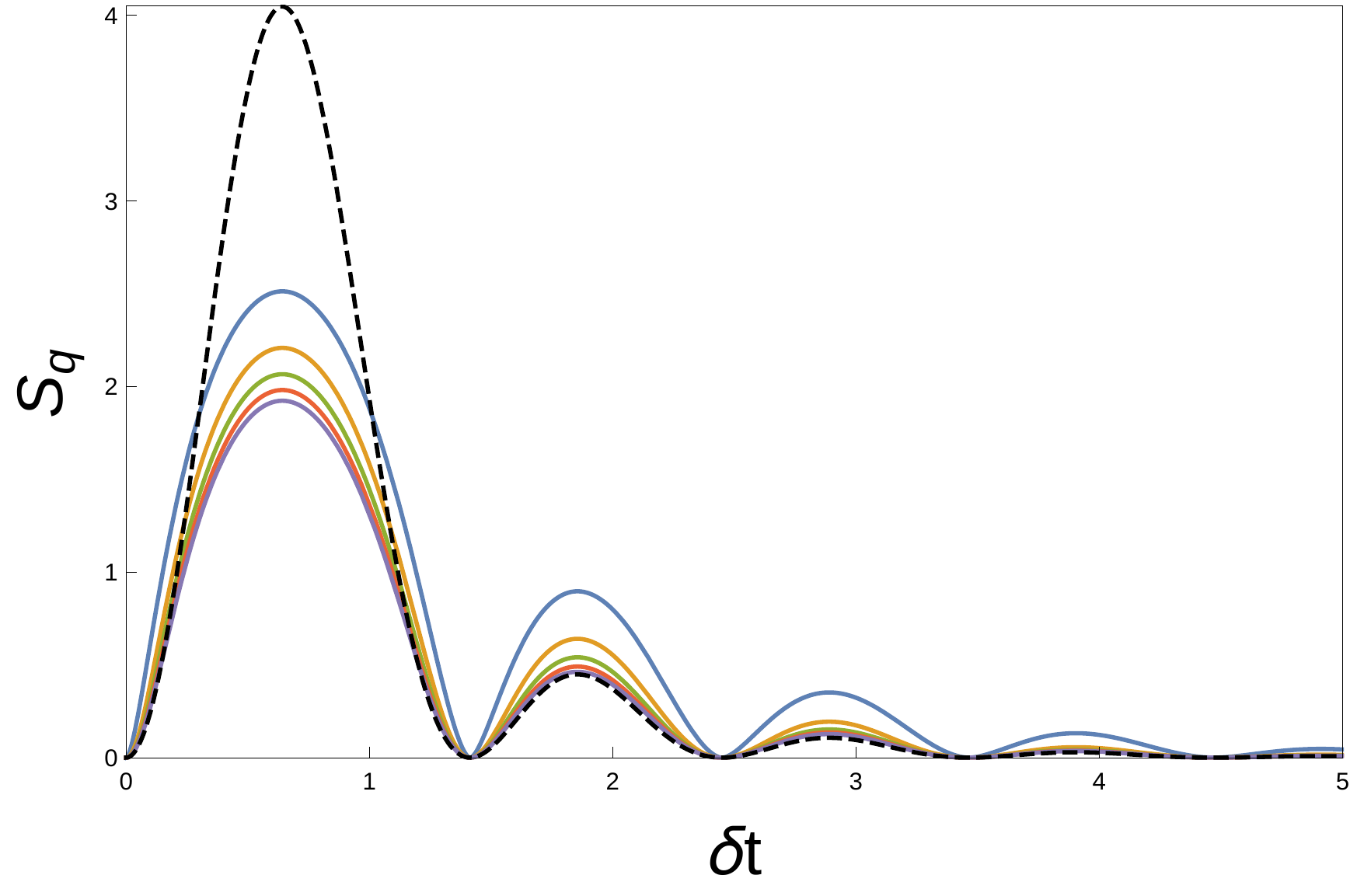}}\quad
    \subfigure[]{\includegraphics[width=.4\textwidth]{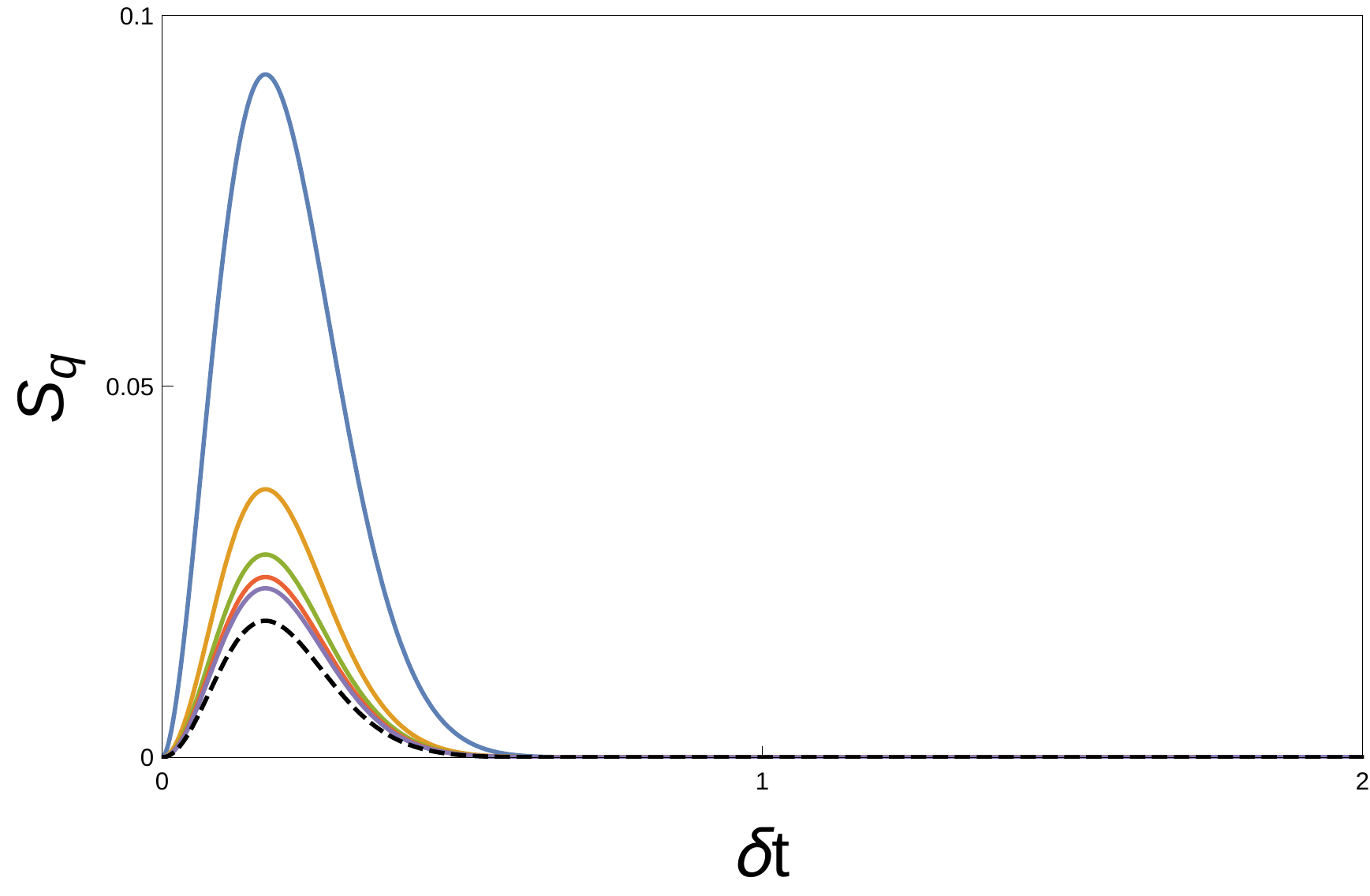}}
\caption{{\small Entanglement entropy $S_\kk$ (blue) and the R\'enyi entropies $S^{(q)}_\kk$ with $q=2,3,4,5$ (from top yellow to bottom purple) as functions of the quench rate $\delta t$. The black dashed curve shows the corresponding particle production rate $|\beta_k|^2$ discussed in \eq{particleproduction}. The mass is fixed to be $m_0=1$. (a) shows the result for low--$k$ modes ($k=0.2$ here), where the oscillatory behavior is more pronounced, while (b) shows the behavior for heavy modes ($k=3$ here) where the oscillations are suppressed and the entropies drop quickly to zero.}}
\label{fig:EERenyivarydtsech}
\end{figure}

\subsection{Entanglement per particle}

It is also interesting to consider the ratio of the entanglement production by the number of particles as a function of $\delta t$, for fixed $k$, and as a function of $k$ for fixed $\delta t$. This is presented in the Figures \fig{EoverNtanh} and \fig{EoverNsech} below for both the Tanh and Sech profiles. It is important to note that this ratio is not constant, besides our earlier remarks that both quantities behave qualitatively similar as a function of the various quench parameters.  

\begin{figure}[htb]
\centering
    \subfigure[]{\includegraphics[width=.4\textwidth]{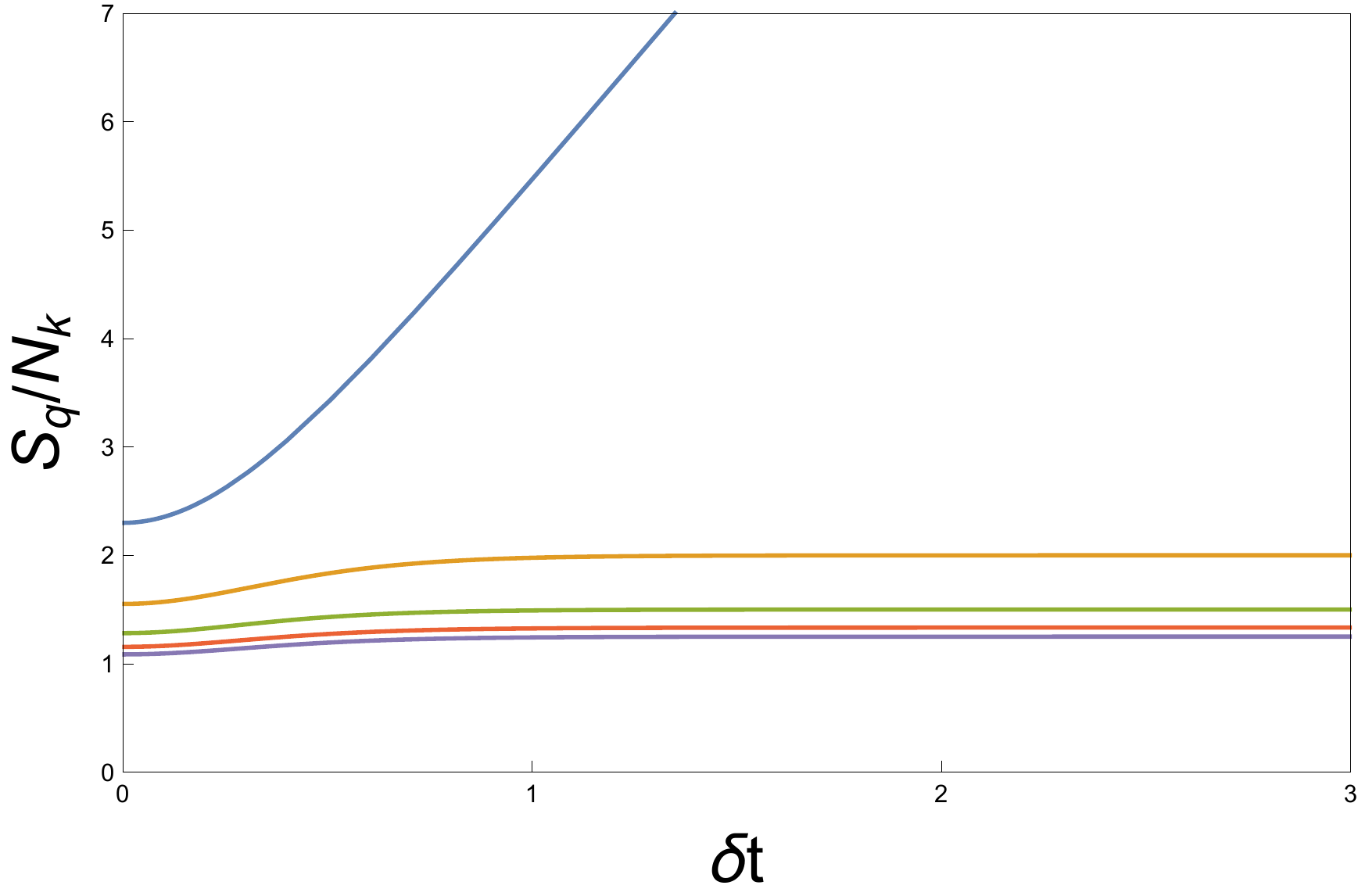}}\quad
    \subfigure[]{\includegraphics[width=.4\textwidth]{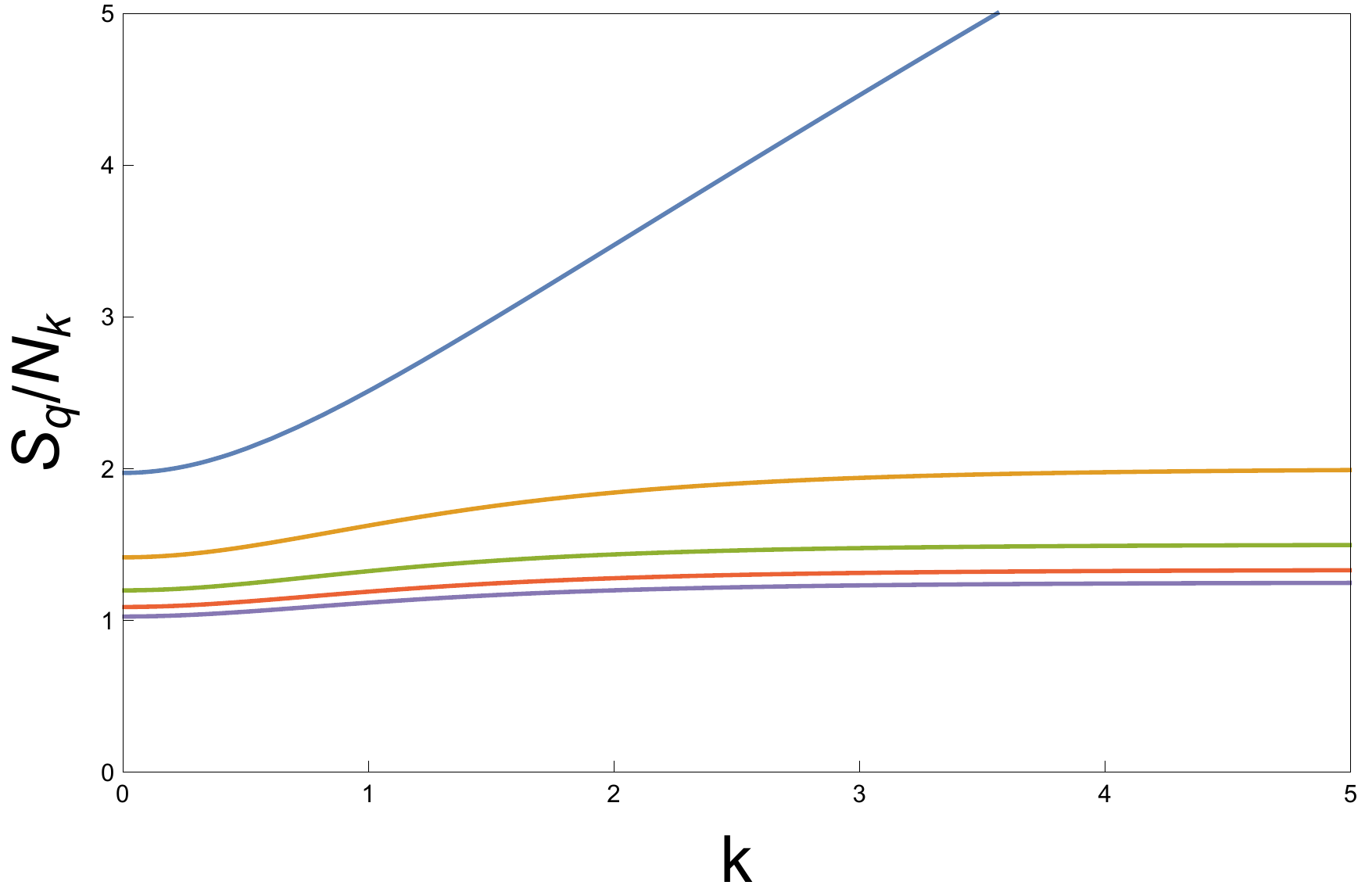}}
\caption{{\small Entanglement per particle produced by the Tanh quench as told by the Entanglement entropy $S_\kk$ (blue) and the R\'enyi entropies $S^{(q)}_\kk$ with $q=2,3,4,5$ (from top yellow to bottom purple) as functions of the quench rate $\delta t$ and of the momentum $k$. The parameters are fixed exactly as in Figure \fig{EERenyivarydt}.}}
\label{fig:EoverNtanh}
\end{figure}

\begin{figure}[htb]
\centering
    \subfigure[]{\includegraphics[width=.4\textwidth]{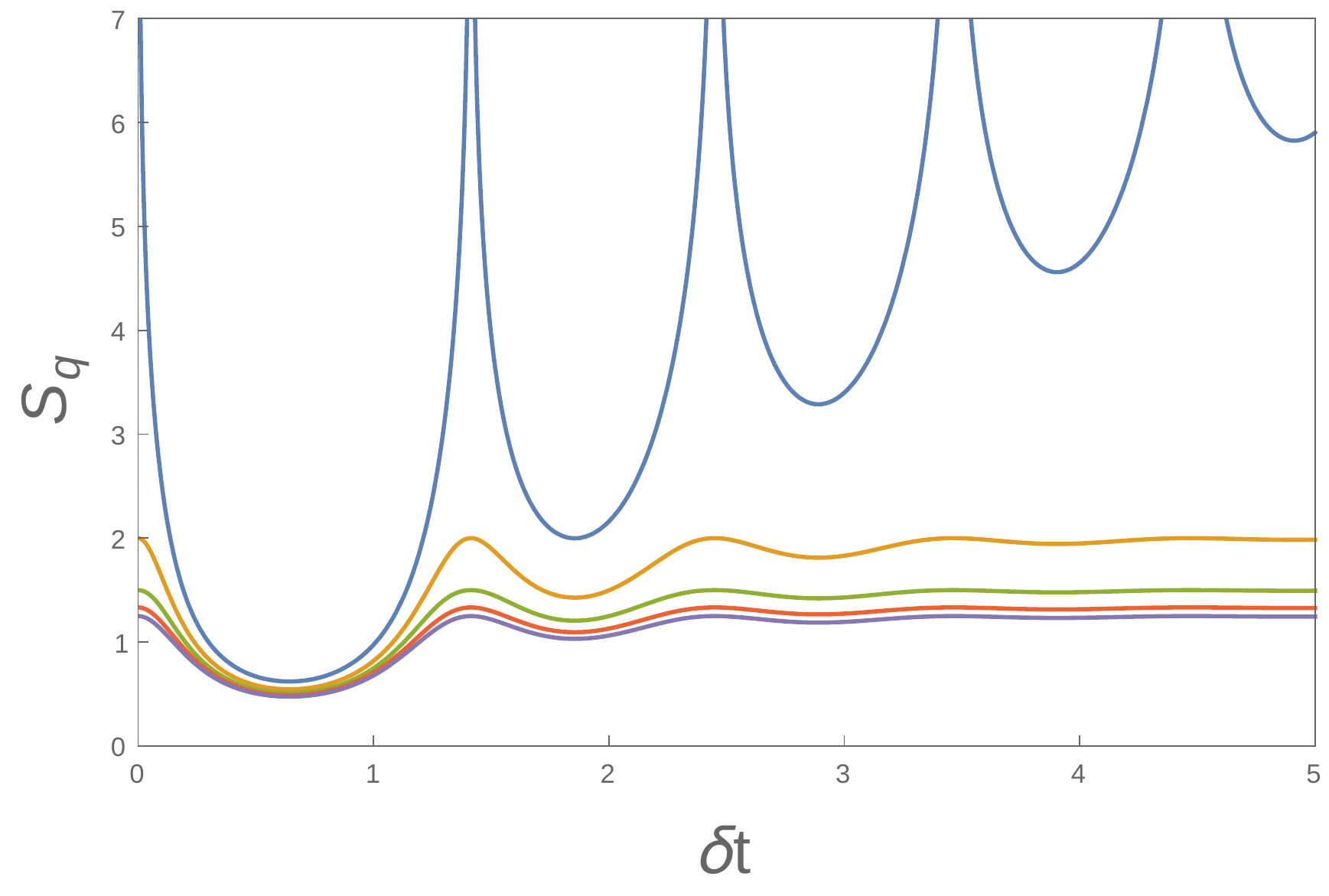}}\quad
    \subfigure[]{\includegraphics[width=.4\textwidth]{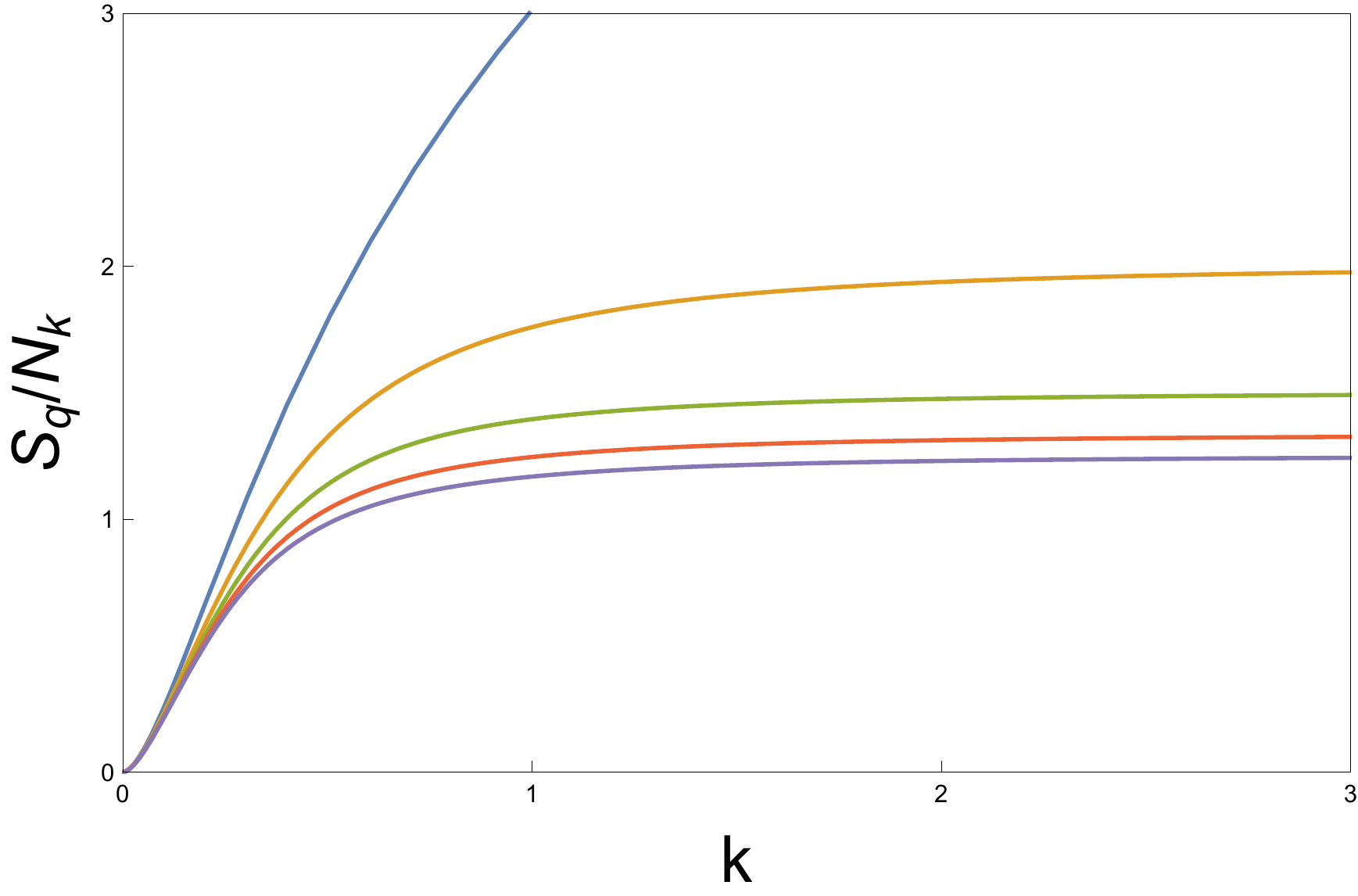}}
\caption{{\small Entanglement per particle produced by the Sech quench as told by the Entanglement entropy $S_\kk$ (blue) and the R\'enyi entropies $S^{(q)}_\kk$ with $q=2,3,4,5$ (from top yellow to bottom purple) as functions of the quench rate $\delta t$ and of the momentum $k$.  The parameters are fixed exactly as in Figure \fig{EERenyivarydtsech} and \fig{EERenyifixeddtsech}.}}
\label{fig:EoverNsech}
\end{figure}

\section{Fermionic case}
\label{sec:fermionic}

A completely analogous construction holds for the case of a mass quench of free Dirac fermions. At the level of equation of motion, the problem amounts to solving the Dirac equation with a time-dependent mass,
\begin{equation}\label{eq:Diraceqquench}
\big[\,\ii\gamma^{\mu}\partial_{\mu}+m(t)\big]\psi=0\,,
\end{equation}
subject to appropriate boundary conditions that allow the identification of positive-frequency modes and asymptotic observers to proceed with the quantization of the model. 
Here, $\gamma^\mu$ are the usual Dirac matrices in $d$-dimensional flat spacetime, defined by the Clifford algebra $\{\gamma^\mu,\gamma^\nu\}=2\eta^{\mu\nu}$. Just as in the bosonic case, the same equation appears when one analyzes a free fermion with constant mass $m$ placed on a curved FLRW spacetime. Specifically, the Dirac equation for such a fermion $\Psi$ minimally coupled to gravity reads
\begin{equation}\label{eq:DiraceqFLRW}
\big[\,\ii\bar{\gamma}^{\mu}(\partial_{\mu}-\Gamma_{\mu})+m\big]\Psi=0\,,
\end{equation}
where $\bar{\gamma}^\mu$ are curved space Dirac matrices (satisfying $\{\bar{\gamma}^\mu,\bar{\gamma}^\nu\}=2g^{\mu\nu}$) and $\Gamma_{\mu}$ the spinorial affine connection associated to the FLRW metric $g_{\mu\nu}$. By performing the conformal rescaling 
\begin{equation}
\Psi\equiv{a(t)}^{\frac{1-d}{2}}\psi
\end{equation}
together with $g_{\mu\nu}=a(t)^2\eta_{\mu\nu}$ we get equation \eq{Diraceqquench} for the transformed fermion $\psi$ that lives in flat spacetime, where the quench profile is related to the cosmological scale factor by
\begin{equation}
 m(t)=m\,a(t)\,.
\end{equation}

Analytical mode solutions to (\ref{eq:Diraceqquench}) have been worked out in \cite{PhysRevD.17.964} for an exactly solvable FLRW model that we now briefly review with appropriate adaptations in order to translate the results for our case of interest, the problem of mass quenches. The model is characterized by the mass profile 
\begin{equation}\label{eq:massquenchtanhfermion}
m\left(t\right)=\frac{1}{2}(m_\text{out}+m_\text{in})+\frac{1}{2}(m_\text{out}-m_\text{in})\tanh\frac{t}{\delta t}\,,
\end{equation}
which is morally the same as \eq{massquenchtanh} in the bosonic case (but notice that here the tanh profile is for $m(t)$, while in that case it was for $m(t)^2$).\footnote{Equivalently, from the curved spacetime point of view the bosonic mass profile \eq{massquenchtanh} corresponds to a FLRW metric with scale factor $a(t)=\sqrt{A+B\tanh\frac{t}{\delta t}}$ while the fermionic one \eq{massquenchtanhfermion} to $a(t)=A+B\tanh\frac{t}{\delta t}$ (with the cosmological parameters $A,B$ related to the ratios $\frac{m_\text{in}}{m},\frac{m_\text{out}}{m}$ as in \eq{minmoutdef}).}
Translational symmetry in the spacelike directions allows the ansatz
\begin{equation}
\psi_{\kk}(t,x)=\big[\gamma^{0}\partial_{t}+\ii\boldsymbol{\gamma}\cdot\kk-m(t)\big]e^{\ii\kk\cdot\x}\phi_{\kk}(t)\,,
\end{equation}
where the function $\phi_\kk(t)$ can be checked to satisfy
\begin{equation}\label{eq:modeeqoscillatorfermion}
 \ddot{\phi}_\kk+\left[\kk^2+m(t)^2+\gamma^0\dot{m}(t)\right]\phi_\kk = 0\,.
\end{equation}
Notice the strong parallel with \eq{modeeqoscillator}. One subtlety, however, is that $\phi_\kk$ here has $d$ components and this is a matrix equation. Writing 
\begin{equation}
 \phi_\kk(t) = \phi_\kk^{(+)}(t)\,v(\boldsymbol{0},\lambda)+\phi_\kk^{(-)}(t)\,u(\boldsymbol{0},\lambda)
\end{equation}
where $v(\boldsymbol{0},\lambda),u(\boldsymbol{0},\lambda)$ are constant zero-momentum eigenspinors of $\gamma^0$ with opposite eigenvalues $\pm\ii$,
\begin{eqnarray}
\gamma^{0}u(\boldsymbol{0},\lambda)&=&-\ii u(\boldsymbol{0},\lambda)\notag\\
\gamma^{0}v(\boldsymbol{0},\lambda)&=&\ii v(\boldsymbol{0},\lambda)\,,
\end{eqnarray}
it follows that the scalar functions $\phi_\kk^{(\pm)}$ also satisfy a harmonic oscillator equation similar to \eq{modeeqoscillator}, but now with an imaginary contribution to the time-dependent oscillator mass. The solutions with \q{in} and \q{out} boundary conditions at $t=\pm\infty$, i.e., behaving as positive-frequency for pre-quench and post-quench observers respectively, are the following
\begin{eqnarray}
 \phi_\kk^{\text{in}\,(\pm)}&=&K_\text{in}\,e^{-\ii\op t-\ii\om\delta t\log\left(2\cosh\frac{t}{\delta t}\right)} {}_2F_{1}\!\left[{\textstyle1+\ii\!\!\left(\om\pm\frac{\delta m}{2}\right)\!\delta t,\ii\!\!\left(\om\mp\frac{\delta m}{2}\right)\!\delta t;1-\oin\delta t;\frac{1+\tanh\frac{t}{\delta t}}{2}}\right]\notag\\
 \phi_\kk^{\text{out}\,(\pm)}&=&K_\text{out}\,e^{-\ii\op t-\ii\om\delta t\log\left(2\cosh\frac{t}{\delta t}\right)} {}_2F_{1}\!\left[{\textstyle1+\ii\!\!\left(\om\pm\frac{\delta m}{2}\right)\!\delta t,\ii\!\!\left(\om\mp\frac{\delta m}{2}\right)\!\delta t;1+\oout\delta t;\frac{1-\tanh\frac{t}{\delta t}}{2}}\right]\notag\\
\end{eqnarray}
where $\oin,\oout$ are the same as previously defined in \eq{omegasdef} and we have introduced 
\begin{equation}
\delta m=m_\text{out}-m_\text{in}\,,\qquad K_\text{in/out}=-\frac{1}{|\kk|}\sqrt{\frac{\omega_\text{in/out}-m_\text{in/out}}{2m_\text{in/out}}}\,.
\end{equation}

The general solution for the fermion $\psi$ can then be written as 
\begin{equation}\label{eq:modeexpansionfermion}
\psi(x)=\frac{1}{(2\pi)^{(d-1)/2}}\int d^{d-1}\kk\,\sqrt{\frac{m_\text{in}}{\oin}}\,\sum_{\lambda=1}^{\lambda_\text{max}}\big[a^\text{in}_{\kk,\lambda}U^\text{in}_{\kk,\lambda}(t,\x)+b^{\text{in}\,{\dagger}}_{\kk,\lambda}V^\text{in}_{\kk,\lambda}(t,\x)\big]\,,
\end{equation}
with the sum over the spinor index running up to $\lambda_\text{max}=2^{d/2-1}$ for $d$ even and $2^{(d-3)/2}$ for $d$ odd, and the curved space spinor mode solutions
\begin{eqnarray}
U^\text{in}_{\kk,\lambda}(t,\x)& \equiv&\left[-\ii\partial_{t}+\ii\kk\cdot\boldsymbol{\gamma}-m(t)\right]\phi_{\kk}^{\text{in},(-)}\,e^{\ii\kk\cdot\x}\,u(\boldsymbol{0},\lambda)\\
V^\text{in}_{\kk,\lambda}(t,\x)& \equiv&\left[\ii\partial_{t}-\ii\kk\cdot\boldsymbol{\gamma}-m(t)\right]\phi_{\kk}^{\text{in},(+)*}\,e^{-\ii\kk\cdot\x}\,v(\boldsymbol{0},\lambda)\,.
\end{eqnarray}
Of course there are analogous expressions for the out modes as well.

A special property of this exactly solvable model, as in the bosonic case, is that the Bogoliubov transformation connecting pre-quench and post-quench modes is diagonal in momentum space. In terms of the functions $\phi_{\kk}^{(\pm)}$ it takes the simple form
\begin{equation}\label{eq:bogofieldfermion}
\phi^{\text{in},(\pm)}_\kk=\alpha_{\kk}^{\pm}\,\phi^{\text{out},(\pm)}_\kk+\beta_{\kk}^{\pm}\,\phi^{\text{out},(\mp)*}_\kk\,
\end{equation}
that allows us to relate the corresponding creation and annihilation operators as
\begin{eqnarray}
a^\text{out}_{\kk,\lambda}&=&\sqrt{\frac{m_\text{in}\,\oout}{m_\text{out}\,\oin}}\,\frac{K_\text{in}}{K_\text{out}}\left(\alpha_{\kk}^{(-)}a^\text{in}_{\kk,\lambda}+\beta_{\kk}^{(-)*}\sum_{\lambda'}X_{\lambda,\lambda'}(-\kk)b^{\text{in}\,\dagger}_{-\kk,\lambda'}\right)\notag\\
b^\text{out}_{\kk,\lambda}&=&\sqrt{\frac{m_\text{in}\,\oout}{m_\text{out}\,\oin}}\,\frac{K_\text{in}}{K_\text{out}}\left(\alpha_{\kk}^{(-)}b^\text{in}_{\kk,\lambda}+\beta_{\kk}^{(-)*}\sum_{\lambda'}X_{\lambda,\lambda'}(-\kk)a^{\text{in}\,\dagger}_{-\kk,\lambda'}\right)\,.
\end{eqnarray}
Here,
\begin{equation}
X_{\lambda,\lambda'}(-\kk)=-2m_\text{out}\,K_\text{out}\,\bar{u}_\text{out}(-\kk,\lambda')\,v(\boldsymbol{0},\lambda)\,
\end{equation}
and $\bar{u}_\text{out}(\kk,\lambda)\equiv \ii u_\text{out}^{\dagger}\left(\boldsymbol{k},\lambda\right)\gamma^{0}$ refers to the polarization $\lambda$ and momentum $\kk$ spinor 
\begin{equation}
u_\text{out}\left(\kk,\lambda\right)= K_\text{out}\left(\ii\gamma^{\mu}k_{\mu}-m\right)u\left(\boldsymbol{0},\lambda\right)\,.
\end{equation}
The Bogoliubov coefficients above are known analytically \cite{PhysRevD.17.964}, namely
\begin{eqnarray}\label{eq:bogocoeffsfermion}
\alpha_{\kk}^{\pm}&=&\frac{\Gamma\left(1-\ii\oin\delta t\right)\Gamma\left(-\ii\oout\delta t\right)}{\Gamma\left(1-\ii\op\delta t\pm \ii\delta m\,\delta t/2\right)\Gamma\left(-\ii\op\delta t\mp \ii\delta m\,\delta t/2\right)}\notag\\
\beta_{\kk}^{\pm}&=&\frac{\Gamma\left(1-\ii\oin\delta t\right)\Gamma\left(\ii\oout\delta t\right)}{\Gamma\left(1+\ii\om\delta t\pm \ii\delta m\,\delta t/2\right)\Gamma\left(\ii\om\delta t\mp \ii \delta m\,\delta t/2\right)}\,.
\end{eqnarray}

With the definitions above we can now proceed as in the previous section and calculate the entanglement production between opposite momentum modes of the fermionic field due the mass quench. We focus on $d=1+1$ dimensions, where the sum over $\lambda$ in \eq{modeexpansionfermion} runs over a single value and considerably simplifies the analysis (also, since the spatial momentum $\kk$ has only one component it can be denoted simply by $k$). 
A similar analysis has been carried out in \cite{Fuentes:2010dt} where the idea was to quantify the entanglement production for a fermionic system due to the cosmological expansion of the FLRW spacetime.

The pre-quench vacuum is defined by 
\begin{equation}
a^\text{in}_{k,\lambda}\,|0_\text{in}\rangle=b^\text{in}_{k,\lambda}\,|0_\text{in}\rangle=0\,
\end{equation}
and the post-quench one $|0_\text{out}\rangle$ can be defined in a similar way. 
The Fock space can again be split into opposite momentum modes $\pm k$, e.g. $|0_\text{in}\rangle=\otimes_k|0_{\text{in}_k}\,0_{\text{in}_{-k}}\rangle$, and since the Bogoliubov transformation \eq{bogofieldfermion} between in and out creation/annihilation operators only mixes $k$ and $-k$ modes one can repeat the steps done previously and express the initial vacuum in terms of the out basis. The result takes the form
\begin{equation}\label{eq:vacuumintooutfermion}
|0_\text{in}\rangle=\bigotimes_{k}\frac{1}{\sqrt{1+|\theta_\mathsmaller{F}|^{2}}}\left(|0_{\text{out}_{k}},0_{\text{out}_{-k}}\rangle-\theta_\mathsmaller{F}\,|1_{\text{out}_{k}},1_{\text{out}_{-k}}\rangle\right)\,,
\end{equation}
with
\begin{equation}
\theta_\mathsmaller{F}\equiv\frac{\beta_{k}^{-*}}{\alpha_{k}^{-*}}\frac{m_\text{out}}{|k|}\left(1-\frac{\oout}{m_\text{out}}\right)\,.
\end{equation}
Here, $|1_{\text{out}_{k}},1_{\text{out}_{-k}}\rangle\equiv a^{\text{out}\,\dagger}_{k,\lambda} b^{\text{out}\,\dagger}_{-k,\lambda}|0\rangle_\text{out}$ denotes the state containing a particle with momentum $k$ and an antiparticle with momentum $-k$ as told by the late time observer. Therefore, we see that the initial vacuum is populated by particle-antiparticle pairs of the out type carrying opposite momenta. This should be contrasted with \eq{vacuumintooutfermion} for bosons, in which case there was an infinite tower of multiparticle excitations thanks to the absence of Pauli's principle in that case.

The density matrix corresponding to the in vacuum state is simply $\rho=|0_\text{in}\rangle\langle0_\text{in}|$ with $|0_\text{in}\rangle$ expressed in terms of the out basis by \eq{vacuumintooutfermion}. 
By focusing on a particular pair ($k,-k$) of momentum modes and tracing out all the antiparticles, we get the very simple reduced state
\begin{equation}\label{eq:rhokfermion}
\rho_{k}\equiv\frac{1}{1+\gf}\left(\,|0_{\text{out}_k}\rangle\langle0_{\text{out}_k}|+\gf\,|1_{\text{out}_k}\rangle\langle1_{\text{out}_k}|\,\right)\,,
\end{equation}
where we have introduced the parameter
\begin{equation}\label{eq:gammafermion}
\gf\equiv |\theta_\mathsmaller{F}|^{2}=\frac{(\om-\delta m/2)(\op-\delta m/2)}{(\om+\delta m/2)(\op+\delta m/2)}\,\frac{\sinh\left[\pi\delta t\left(\om+\delta m/2\right)\right]\sinh\left[\pi\delta t\left(\om-\delta m/2\right)\right]}{\sinh\left[\pi\delta t\left(\op+\delta m/2\right)\right]\sinh\left[\pi\delta t\left(\op-\delta m/2\right)\right]}\,
\end{equation}
and used \eq{bogocoeffsfermion} to simplify it. This is the fermionic analog of $\gb$ in the bosonic case (see \eq{gamma}).

At this point the technical difficulties of solving the Dirac equation with a time-dependent mass faced above really pay off when it comes to calculating the entanglement properties: the reduced state \eq{rhokfermion} is just a two-dimensional diagonal matrix (compare with the infinite sum over occupation numbers in \eq{rhok}). The R\'enyi entropies of order $q$ are immediately found to be
\begin{equation}\label{eq:Renyifermion}
S_k^{(q)}\equiv\frac{1}{1-q}\log\left[\frac{1+{\gf}^{q}}{\left(1+\gf\right)^{q}}\right]\,,
\end{equation}
and taking the limit $q\to1$ we obtain the entanglement entropy 
\begin{equation}\label{eq:EEfermion}
S_k = \lim_{q\to1} S_k^{(q)} = \log\frac{1+\gf}{\gf^{\gf/(\gf+1)}}\,.
\end{equation}
which agrees with the expression originally obtained in \cite{Fuentes:2010dt} in the framework of an expanding spacetime. Notice that the result for the entropies would have been the same had we traced out the particles instead of the antiparticles in the beginning, since the total state is pure.

The total amount of entanglement produced by the quench at late times is measured by these entropies and depends on the masses $m_\text{in},m_\text{out}$, the mode $k$, and the time scale $\delta t$. Similarly to the free boson case, it is fair to state that the parameter $\gf$ encodes all the the late-time entanglement properties between opposite momentum modes. The resemblance between the formulas \eq{Renyifermion},\eq{EEfermion} and \eq{Renyi},\eq{EE} for the fermionic and bosonic entropy production is striking. In fact, the expressions for the Renyi entropies can be converted into minus one another under $\gf\leftrightarrow-\gb$. Notice that this simple relation between the bosonic and fermionic R\'enyi entropies does not commute with the limit $q\to1$, i.e., it is not shared by the expressions for the EE. 

The entropies are again monotonic functions of $\gf$ and can be inverted to determine $\gf(S)$, i.e., the information concerning mode entanglement is in principle enough to determine all the quench parameters. An important difference with respect to the bosonic result, however, is that the EE in the present case is limited from above by $\log2\approx0.7$, since the reduced state for particles with momentum $k$ lives in a two-dimensional Hilbert space. In particular, even when one of the masses vanishes (i.e., when the quench crosses a critical point) this upper bound prevents the occurrence of the divergent zero mode contribution that takes place for bosons. Another important difference is that now the expressions are not symmetric under $\oin\leftrightarrow\oout$, so the behavior for $m_\text{in}>m_\text{out}$ is distinct from $m_\text{in}<m_\text{out}$ (or $\delta m>0$ and $<0$, respectively) and must be analyzed separately, as we shall see.

Figure \fig{EERenyifermionfixeddt} shows the $k$-dependence of the EE and the second Renyi entropy for a fixed quench rate $\delta t$. The lefthand side figure corresponds to quenches that decrease the mass, while the other two to increasing-mass quenches. For the former the entropies are always monotonically decreasing functions of $k$, showing that entanglement production is more noticeable between IR modes. However, for the case of increasing-mass quenches this is only true up to a critical value of $\delta m$ (see (b) and (c)). Above this critical value of $\delta m$ this monotonic behavior is broken as shown in the right figure. Interestingly, this indicates that for large enough final masses the maximal entanglement production is not achieved at IR modes but rather at an intermediate momentum value $k=k_\text{max}$. 

Figure \fig{EERenyifermionvarydt} illustrates the dependence of the entropies on the time scale $\delta t$. Cases (a), (b), (c) show the result for a single mode $k$ in a quench with $\delta m<0$ (the former) and $\delta m>0$ (the latter two), respectively, while (d), (e), (f) show the total entanglement produced in each of the two situations, obtained after integration over all $k\ge0$ (this is the left-right entropy studied, e.g., in \cite{PandoZayas:2014wsa}). We see that in mass-decreasing quenches the entanglement production is always bigger the faster the quench is done. For a single mode $k$, this is still true for mass-increasing quenches up to a limiting value $\delta m^*>0$ ((a) and (b)) but fails to be true for deformations stronger than this (case (c)), where maximum production is then achieved at an intermediate value of $\delta t$. This unusual feature disappears when one integrates over all modes $k$, as shown in (d). Notice also that again adiabatic quenches do not produce any mode entanglement, since all curves asymptote to zero as $\delta t$ grows. 

\begin{figure}[ht]
\begin{center}
    \subfigure[$\delta m\le0$]{\includegraphics[width=.32\textwidth]{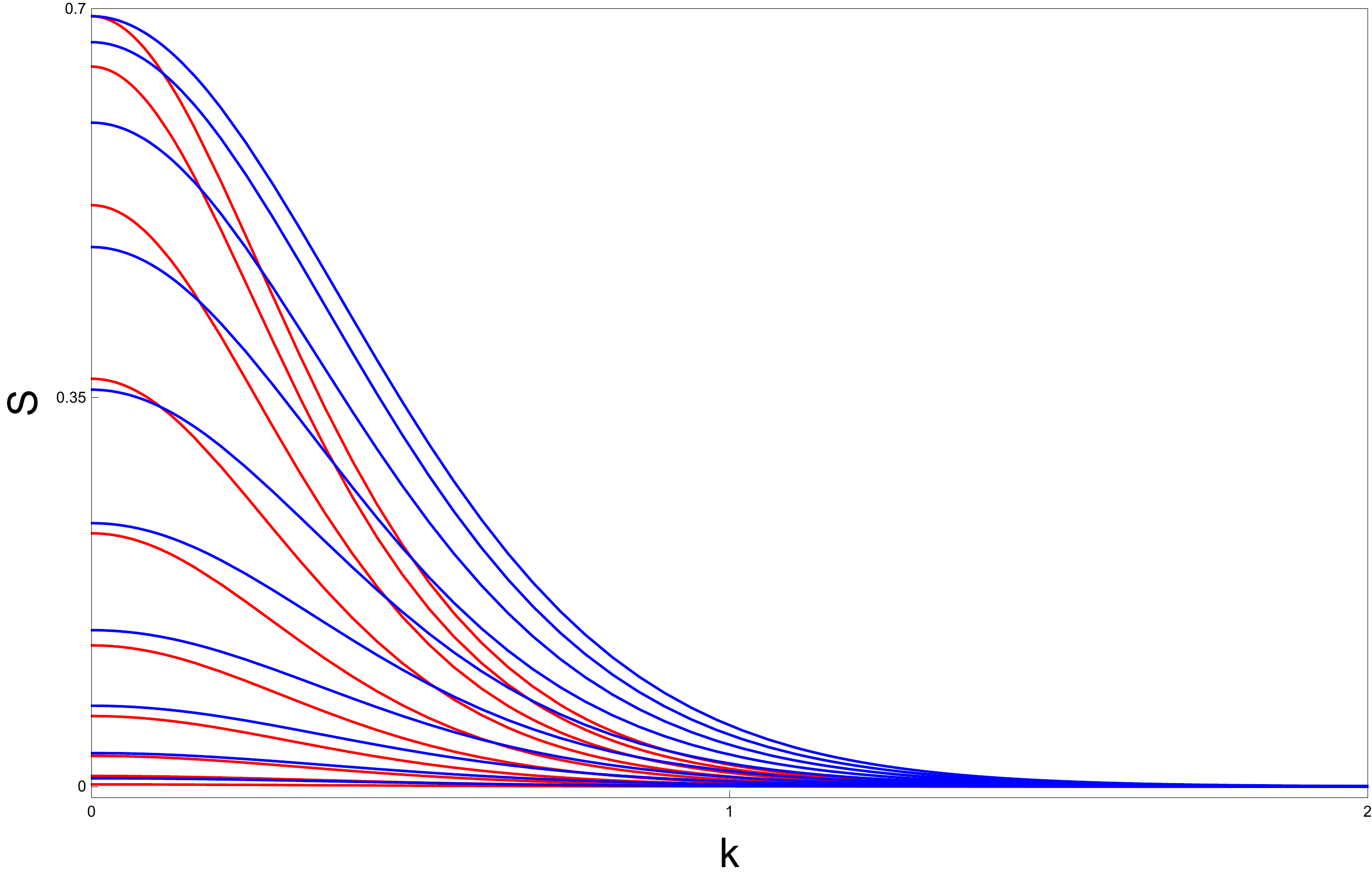}}
    \subfigure[$0<\delta m<\delta m^*$]{\includegraphics[width=.32\textwidth]{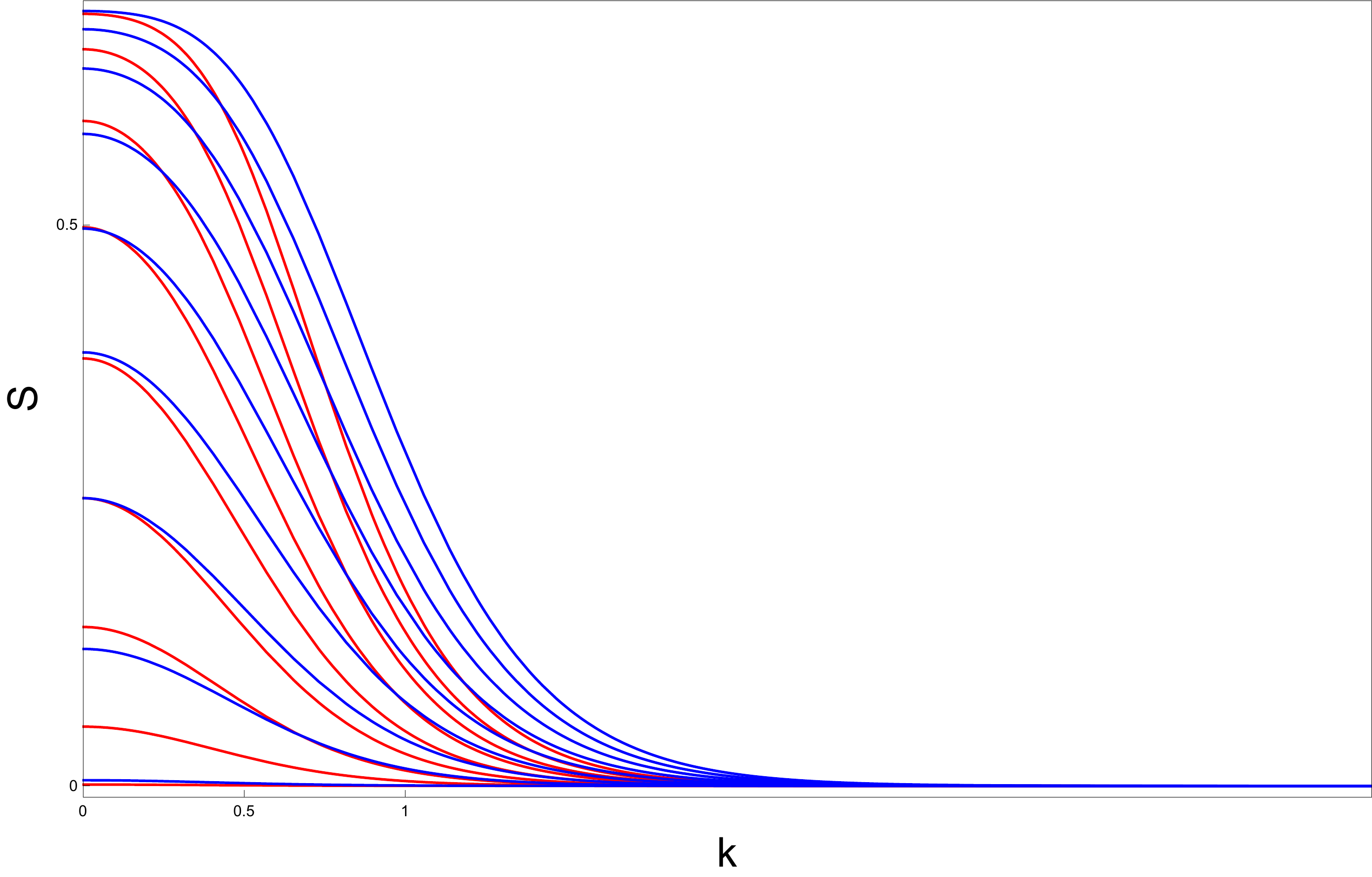}}
    \subfigure[$\delta m>\delta m^*	$]{\includegraphics[width=.32\textwidth]{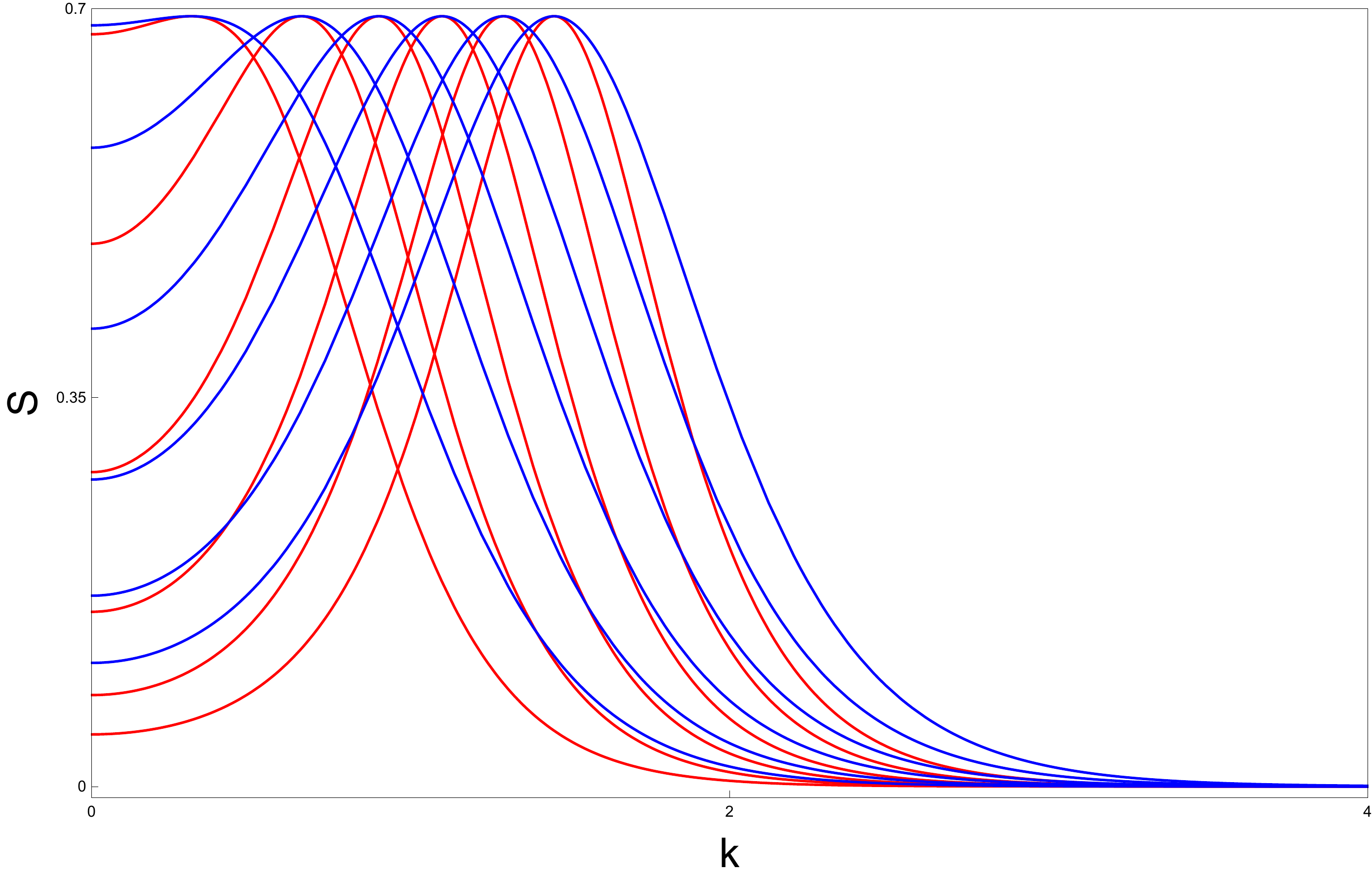}}
\end{center}
\caption{{\small $k$-dependence of the entanglement entropy $S_k$ (blue) and the second Renyi entropy $S^{(2)}_k$ (red) at a fixed quench rate $\delta t=1$ for fermionic mass quenches. (a) corresponds to quenches that decrease the mass, while (b) and (c) to increasing mass. In the plots we have set $m_\text{in}=1$ and selected different values of $m_\text{out}$ (equivalently $\delta m$). In (a), the values of $m_\text{out}$ grow from zero (top) to $m_\text{in}$ (bottom); in (b), the values of $m_\text{out}$ grow from $m_\text{in}$ (bottom) up to a critical value $m_\text{out}^*$ (top) where the functions stop being monotonic ($m_\text{out}^*\approx 10$ for the choice of parameters mentioned above); then, in (c), $m_\text{out}$ keeps increasing above this critical value as the curves go from left to right.}}
\label{fig:EERenyifermionfixeddt}
\end{figure}

\begin{figure}[ht]
    \subfigure[$\delta m\le0$]{\includegraphics[width=.32\textwidth]{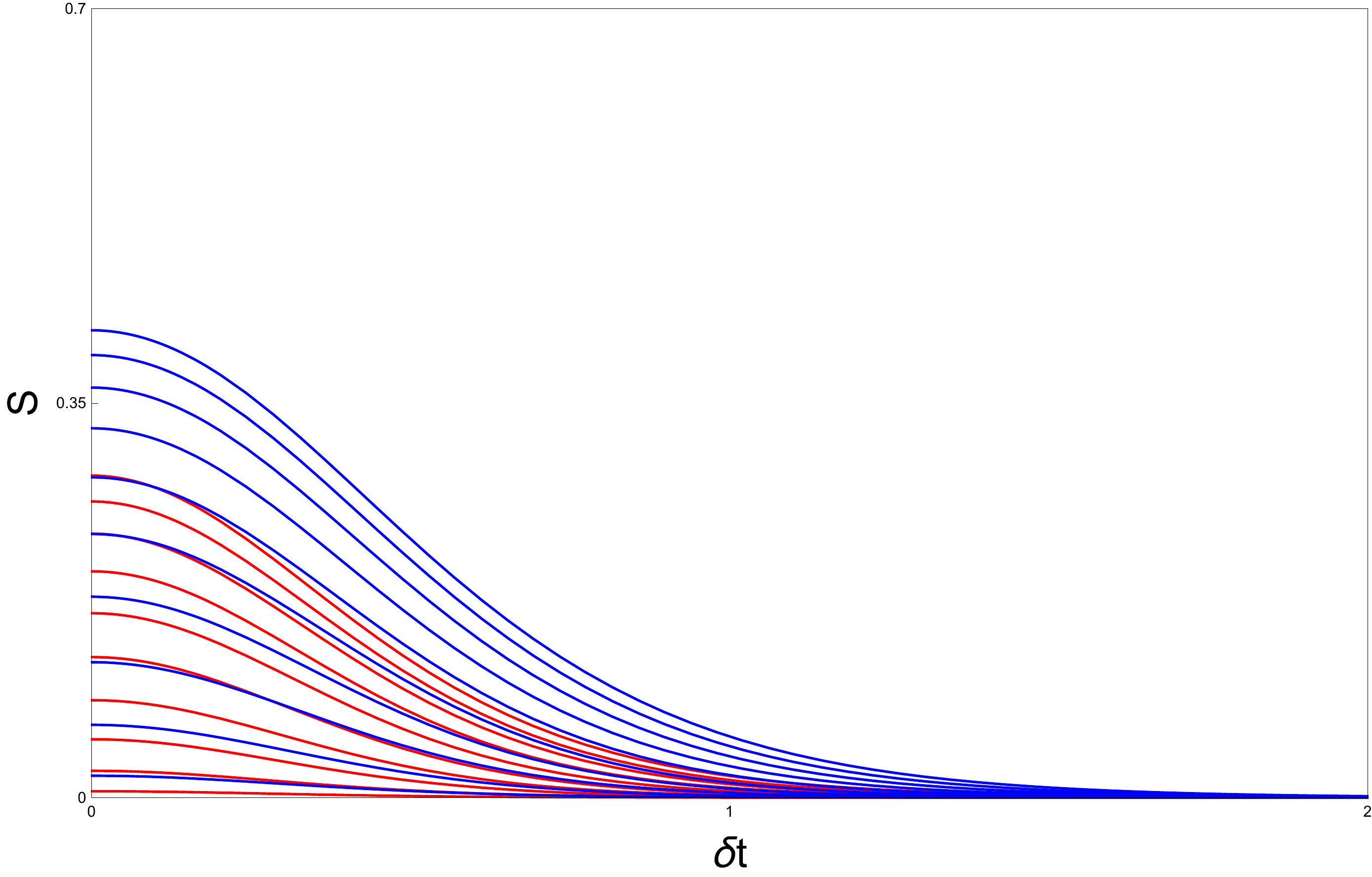}}
    \subfigure[$0<\delta m<\delta m^*$]{\includegraphics[width=.32\textwidth]{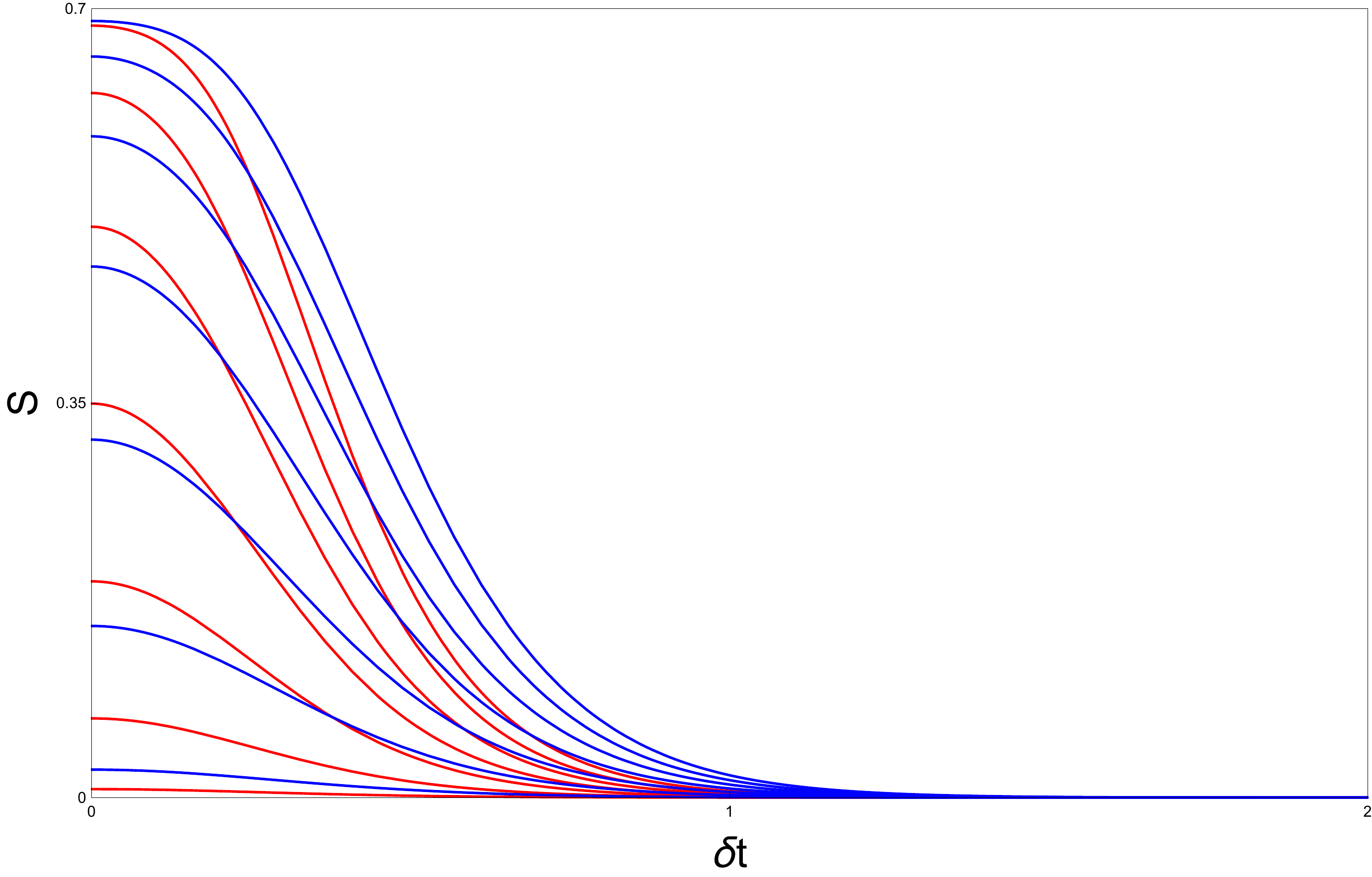}}
    \subfigure[$\delta m\ge\delta m^*$]{\includegraphics[width=.32\textwidth]{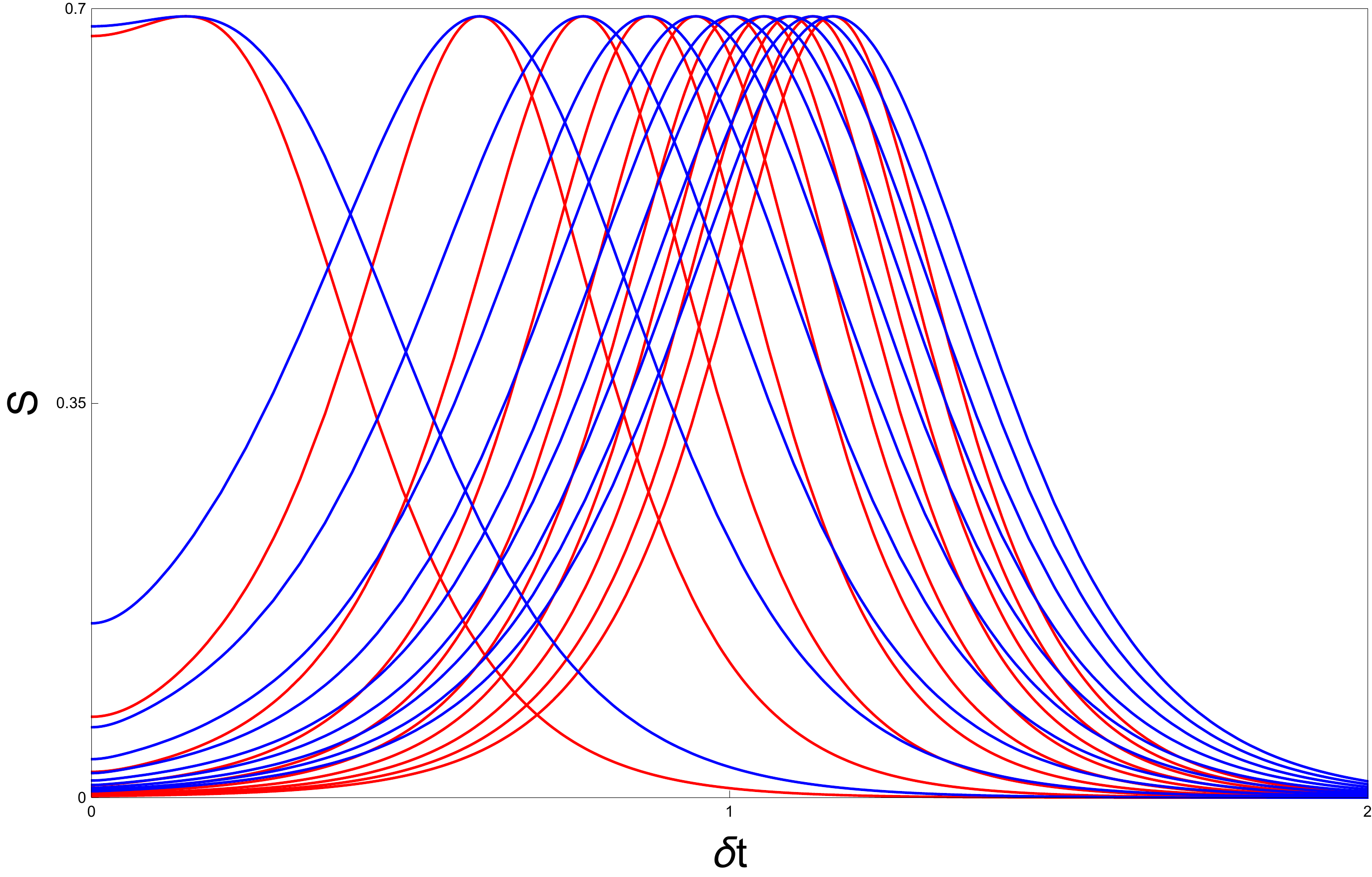}}
    \\
    \subfigure[$\delta m\le0$]{\includegraphics[width=.32\textwidth]{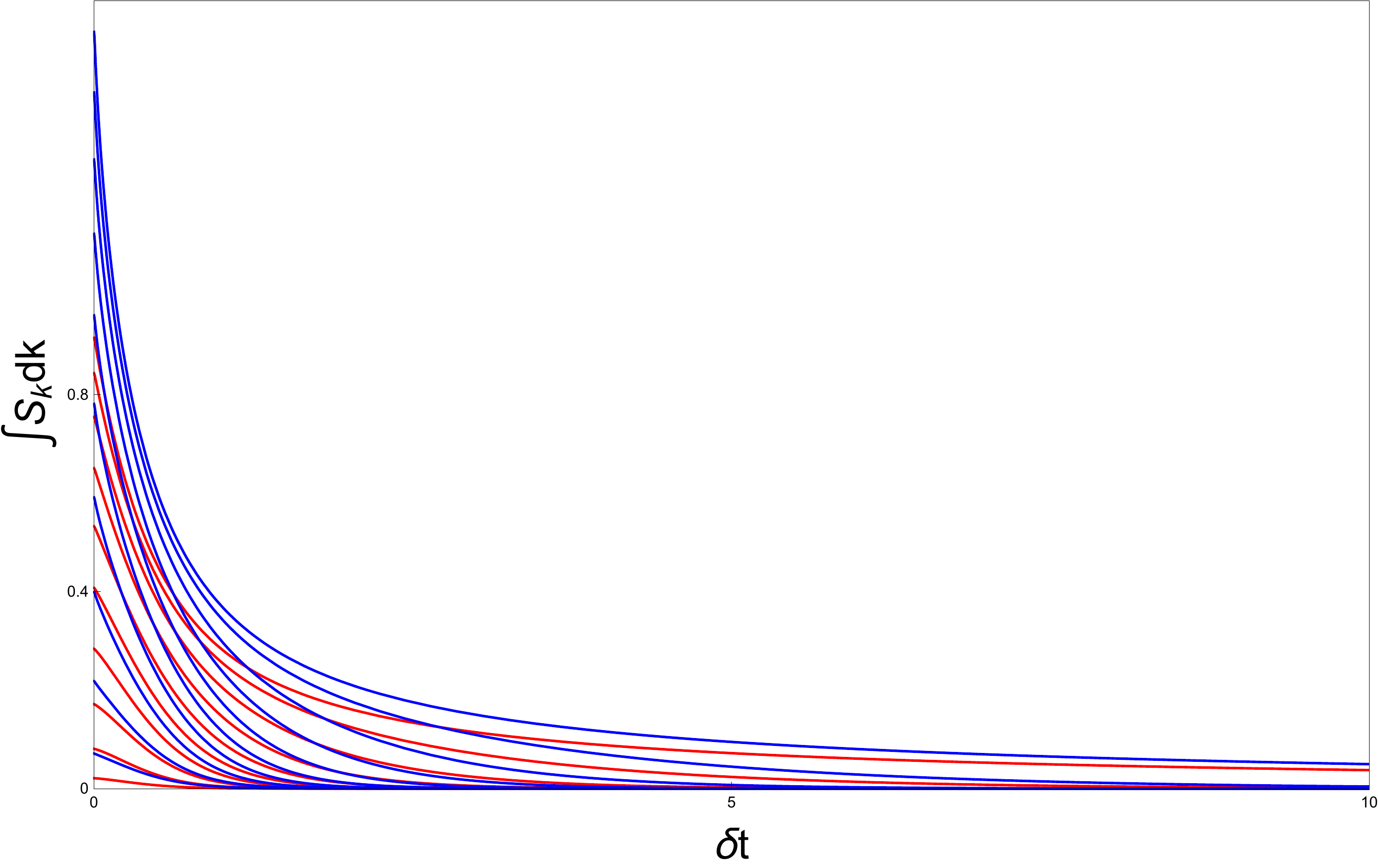}}
    \subfigure[$0<\delta m<\delta m^*$]{\includegraphics[width=.32\textwidth]{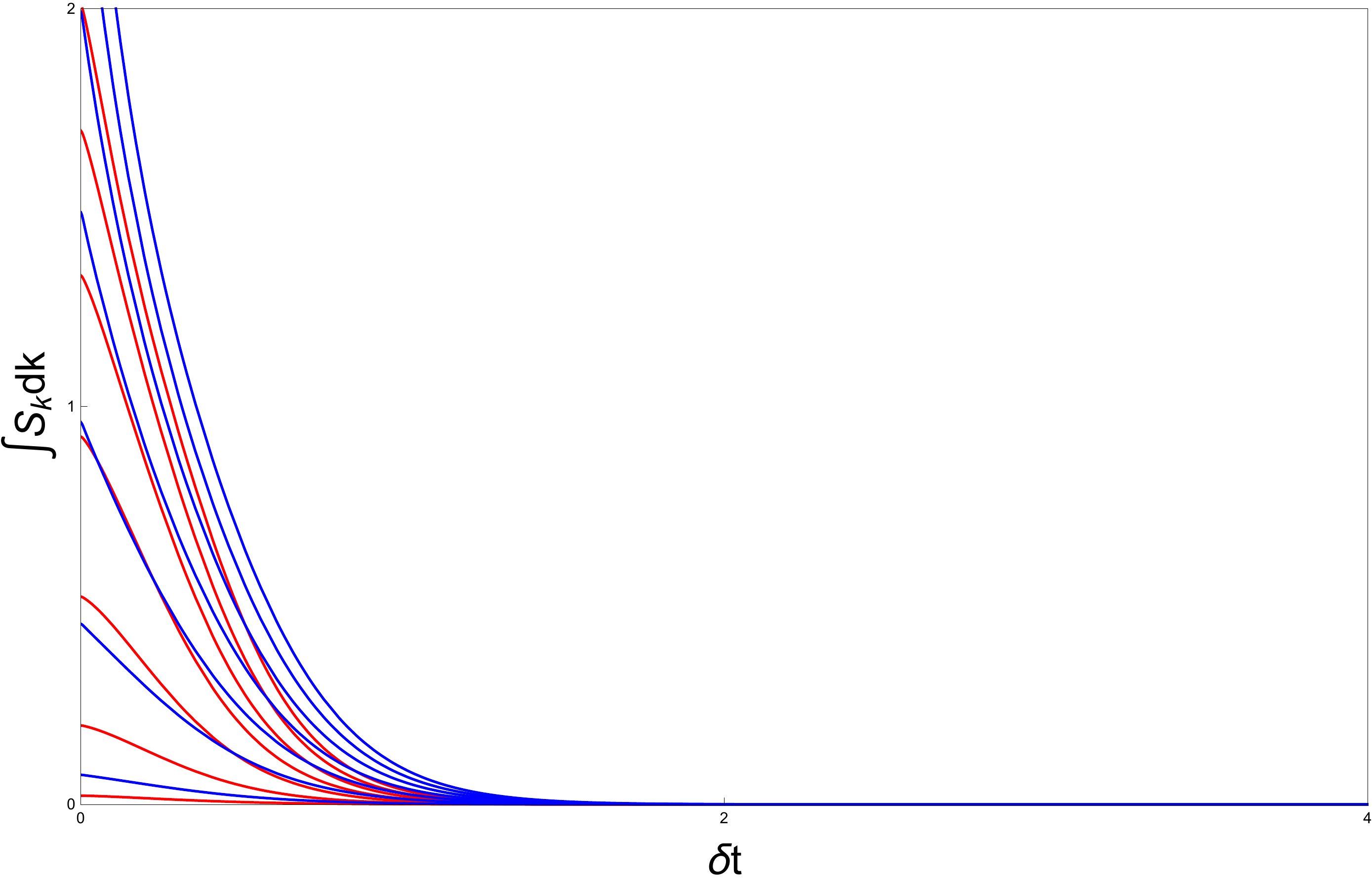}}
    \subfigure[$\delta m\ge\delta m^*$]{\includegraphics[width=.32\textwidth]{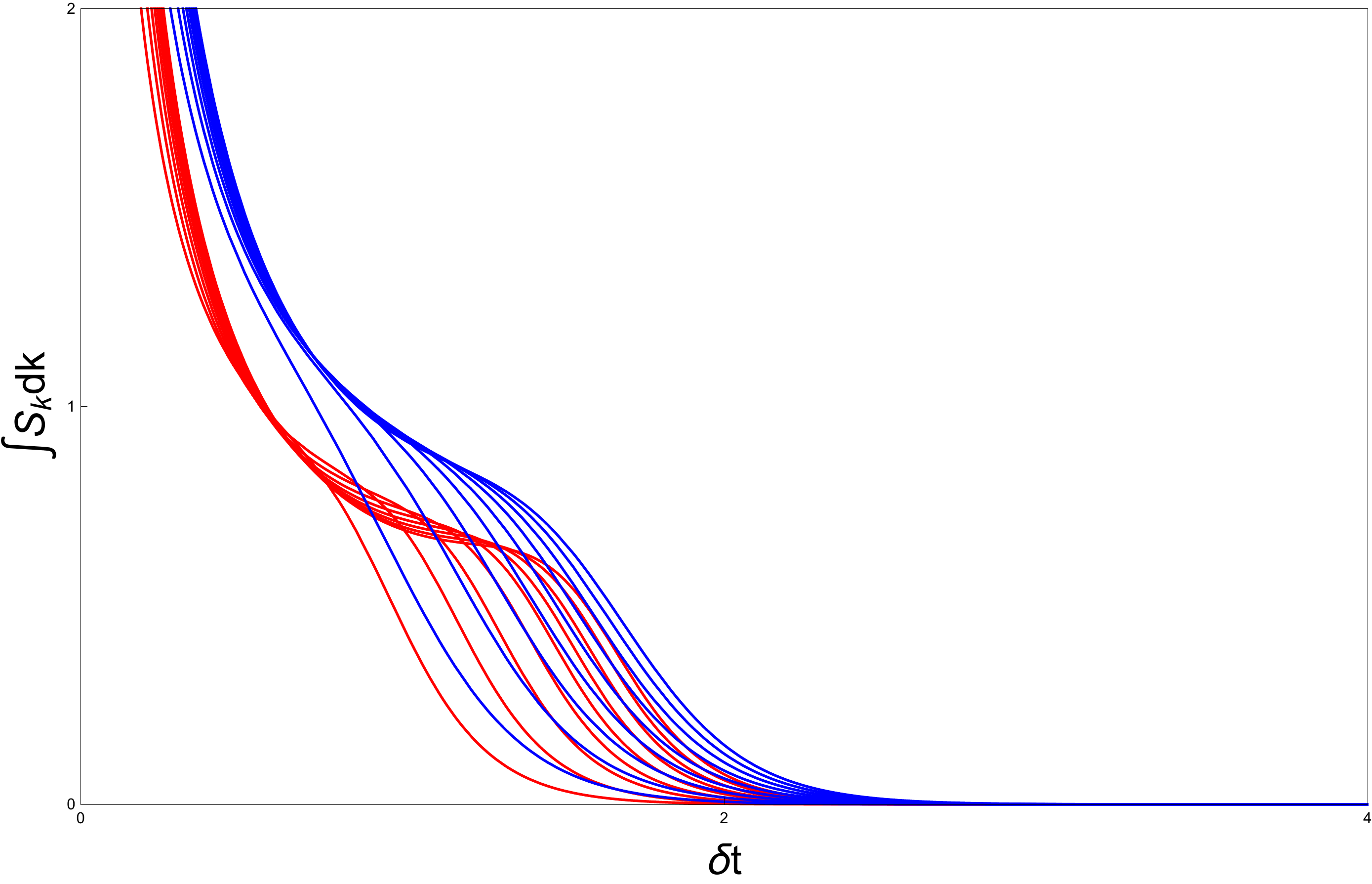}}
\caption{{\small Entanglement entropy $S_k$ (blue) and the second Renyi entropy $S^{(2)}_k$ (red) as functions of the quench rate $\delta t$ for a particular mode $k=1$. (a) shows the result for quenches that decrease the mass, while (b) and (c) for those that increase the mass. For the plots we have set $m_\text{in}=1$ and chosen different values of $m_\text{out}$ in the same way as explained in Figure \fig{EERenyifermionfixeddt}. The critical value separating the last two behaviors is $m_\text{out}^*\approx 2.4$ for the choice of parameters mentioned above. We also show in (d), (e), and (f) the total left-right entropies obtained by integrating (a), (b), and (c) over all momenta $k\in[0,\infty)$, respectively.}}
\label{fig:EERenyifermionvarydt}
\end{figure}

\section{Connection with the Generalized Gibbs Ensemble}
\label{sec:GGE}

In this section, we will show that our above-mentioned results for the Renyi and entanglement entropies agree with the predictions of a Generalized Gibbs Ensemble, showing that this steady state correctly describes the late time dynamics after smooth quenches even from the point of view of entanglement measures (in addition to the well-understood case of correlation functions). 

Recall that while most quantum systems are known to thermalize in the usual sense of approaching a Gibbs ensemble at late times, integrable systems such as our free field models thermalize in the more general sense of a Generalized Gibbs Ensemble (GGE). The density matrix for a GGE state is given by \cite{2007PhRvL..98e0405R}
\begin{equation}\label{eq:GGEdef}
\GGE=Z^{-1}\,e^{-\sum_{j}\lambda_{j}\,I_{j}}\,
\end{equation}
where $\{I_j\}$ denotes a full set of conserved charges, $Z=\Tr\left(e^{-\sum_{j}\lambda_{j}\,I_{j}}\right)$ is the partition function, and $\{\lambda_j\}$ are Lagrange multipliers, fixed by demanding that the GGE value of the conserved charges coincides with their initial value,
\begin{equation}\label{eq:lambdadef}
 \Tr\left(\GGE\,I_j\right) = \langle I_j\rangle_\text{initial}\,.
\end{equation}
The precise statement about the equilibration of integrable systems, therefore, is that expectation values of local operators in the model converge to their GGE ensemble values at late times or, equivalently, that the reduced state of any finite subsystem approaches the corresponding reduced density matrix of GGE.

The question remains of which integrals of motion should appear in the GGE for a given model. For free boson theories in $1+1$ dimensions and Gaussian initial states (such as the vacuum of quadratic Hamiltonians), it was shown in \cite{Sotiriadis:2014uza} that a good choice for these operators is the particle number $N_k=a^\dagger_k a_k$ at each momentum mode (the $\sum_j$ then becomes $\int dk$). In our case of interest \eq{rhototal}, where we deal with a single pair of modes ($k,-k$) in the out-basis, after tracing over all the other irrelevant momenta to get a bunch of unit factors the GGE is simply
\begin{equation}
 \GGE=\frac{1}{Z}e^{-\lambda^{}_{k}N^\text{out}_{k}-\lambda^{}_{-k}N^\text{out}_{-k}}\,,
\end{equation}
where the partition function is
\begin{equation}
Z=\Tr\left(e^{-\lambda^{}_{k}N^\text{out}_{k}-\lambda^{}_{-k}N^\text{out}_{-k}}\right)=\frac{1}{\left(1-e^{-\lambda_{k}}\right)\left(1-e^{-\lambda_{-k}}\right)}\,.
\end{equation} 

After tracing out the negative-momentum modes one is left with the reduced GGE density matrix
\begin{equation}\label{eq:GGEreduced}
{\GGE}_k = \left(1-e^{-\lambda_{k}}\right)\,e^{-\lambda_{k}N^\text{out}_{k}}
\end{equation} 
describing the positive ones. The lagrange multipliers $\lambda_k$ are defined by the condition \eq{lambdadef}, namely $\Tr\left({\GGE}^{}_k N^\text{out}_{k}\right)=\langle0_\text{in}|N^\text{out}_{k}|0_\text{in}\rangle \equiv n_k$, which gives
\begin{equation}
\lambda_{k}=\log\frac{n_{k}+1}{n_{k}}\,.
\end{equation}
Now it is straightforward to check (see \eq{particleproduction} for a similar calculation) that the occupation number $n_k$ is determined by the Bogoliubov coefficient $\beta_k$ as $n_k=|\beta_k|^2$, and using the algebraic relation $|\beta_{k}|^{2}+1=|\alpha_{k}|^{2}$ (see \eq{bogocoeffssquared}) we get
\begin{equation}\label{eq:lambdagamma}
 \lambda_k = -\log\gb\,,
\end{equation}
where $\gb$ is defined by \eq{gamma}. This is a remarkable result which is worth emphasizing: the Lagrange multipliers that characterize the GGE are completely determined by the quench parameters $\mins,\mouts,\delta t$ and the momentum mode $k$ via the parameter $\gb$. This gives support to the results of \cite{Mandal:2015kxi}, where thermalization to a GGE was argued to hold at the level of two-point correlators.

Indeed, if we calculate the Renyi entropy associated to the reduced GGE state \eq{GGEreduced} we obtain
\begin{equation}
S_{k}^{\left(q\right)}=\frac{1}{1-q}\log\left(\Tr{{\GGE}_{k}}^{q}\right)=\frac{1}{1-q}\log\frac{\left(1-e^{-\lambda_{k}}\right)^{q}}{1-e^{-\lambda_{k}q}}\,,
\end{equation} 
which is in perfect agreement with our result \eq{Renyi} after identifying the parameters according to \eq{lambdagamma}. 
This shows that the GGE prediction for the late-time dynamics of free bosons as proposed in \cite{Sotiriadis:2014uza} indeed works for our model (as told by entanglement measures). 

A similar construction should hold also for the fermionic model, in which case the Lagrange multipliers characterizing the GGE would be fully determined by the fermionic parameter $\gf$ of \eq{gammafermion}. 

\section{Conclusions}
\label{sec:conclusions}
\indent

In this work, we have calculated the late-time entanglement in momentum space produced after smooth mass quenches for free scalar (in $d$ dimensions) and fermion theories (in 1+1 dimensions). 
The strategy is inspired by \cite{Ball:2005xa,Fuentes:2010dt}, which studied the entanglement production for quantum fields in an expanding universe, and can be translated to the problem of a quench through a Weyl rescaling of the field. 
The initial state was taken to be the vacuum as defined by a pre-quench observer. As the quench is performed, particle excitations carrying all possible momentum modes are produced as told by a post-quench observer. We then calculated the entanglement and R\'enyi entropies between a single quantum field mode $\kk$ and its opposite mode carrying momentum $-\kk$, which is the only non-trivial case in our exactly solvable model (i.e., no entanglement is produced between other pairs ($\kk,\kk'$) of modes other than this). The results have simple analytical formulas and are expressed in terms of a single parameter, $\gb$ for bosons and $\gf$ for fermions, that encodes all the late-time entanglement properties. 

For ($1+1$)-dimensional bosons, we have shown that our results match the predictions of a Generalized Gibbs Ensemble where the conserved charges are taken to be the mode number operators as defined by the post-quench observer, namely $N^\text{out}_{k}=a^{\dagger\,\text{out}}_{k}a^\text{out}_{k}$ (for all $k$). We were able to precisely calculate the Lagrange multipliers that characterize the GGE in terms of the parameters $m_\text{in},m_\text{out},\delta t$ defining the quench, namely $\lambda_k=-\log\gb$. Hence, having fully specified the GGE state, the long-time behavior of any local observable after the quench can now be calculated (not only the entanglement and Renyi entropies presented here). 

In the bosonic case, for both the Tanh and the Sech mass profiles, we saw that at any fixed quench rate $\delta t$ more entanglement is produced between light modes (the ones carrying small momentum $k$) than between heavy ones, as expected. The entanglement production is monotonically decreasing with $k$ and its magnitude grows with the magnitude of the quench, characterized by the absolute value of the difference $\delta m^2=\mouts-\mins$ of the initial and final mass in Tanh case and by $m_{0}^2$ (the maximum value of the mass during the quench) in the Sech case. The picture is qualitatively similar to the particle production rate given by $n_k=|\beta_k|^2$. 

The dependence on $\delta t$ for a given mode $\kk$ is more interesting: for both profiles it is true that more entanglement is produced for faster quenches (small $\delta t$) with respect to slow ones, while adiabatic quenches ($\delta t\to\infty$) do not produce entanglement at all. In particular, this means that, among all the quenches reaching some $\mouts\ne0$, maximal entropy production is achieved for the one that started from a CFT (i.e., $\mins=0$). However, for the Tanh profile the entanglement is found to decrease monotonically as a function of $\delta t$, while for the Sech profile the decrease is non-monotonic, being given by damped oscillations modulated by $k$.

In the fermionic case the results are more subtle. First we noted that the sign of $\delta m=m_\text{out}-m_\text{in}$ becomes important, that is, the result here is not invariant under $m_\text{in}\leftrightarrow m_\text{out}$ and, as a consequence, whether the quench increases or decreases the mass has a significant impact on the final result. For a fixed rate $\delta t$, and for a quench decreasing the mass, we find a very similar result to that of the Tanh profile for bosons. For quenches that increase the mass, however, the results remain similar to that of bosons only up to a certain critical value $\delta m^{*}$, above which the entanglement ceases to be a monotonic funtion of $|k|$. Instead, as we increase the mass beyond this critical value, we observe a peak at some intermediate value $k_\text{max}$ that grows as $\delta m$ grows. In other words, for quenches that increase too much the mass it fails to be true the statement that more entanglement is produced between light modes. 
Also, for a particular mode $k$, the behaviour of the entanglement and R\'enyi entropies as a function of $\delta t$ has the same features discussed in the last paragraph. Namely, for quenches that decrease the mass the result is very similar to the bosonic one, while for quenches that increase the mass this holds true only up to a critical value $\delta m^{*}$ above which the entropies cease to decrease monotonically with $\delta t$, becoming instead peaked at an intermediate value. If we integrate over all modes $k$ to get the total entanglement production the entropies remain monotonically decreasing, although for $\delta m > \delta m^{*}$ we can still notice a small bump at an intermediate value of $\delta t$.  

We believe these results shall hold (at least qualitatively) for generic mass quenches in free field theories, that is, regardless of the precise form of the quench profile $m(t)$. For instance, we expect the same behavior as in the Tanh case for any other quench profile that goes from the same $m_\text{in}$ to the same $m_\text{out}$ within a finite scale $\delta t$. This expectation is supported even by comparing the qualitative results for the Tanh and the Sech profiles (which do not share the same initial and final masses), namely the fact that the entanglement production is more significant for light modes and faster quenches. Of course we have no formal proof of that, but it would be very surprising if by simply changing $m(t)$ the $k$ or $\delta t$ dependence of the entropies happened to be qualitatively different.

As future prospects, one would like to perform similar calculations for weakly coupled quantum field theories in order to access how the presence of interactions modify the results. 
Another interesting generalization would be a geometric prescription for computing momentum-space entanglement in AdS/CFT, by generalizing the Ryu-Takayanagi prescription \cite{Ryu:2006bv,Hubeny:2007xt}) which is appropriate to deal only with real-space entanglement entropy. This was already speculated on \cite{Balasubramanian:2011ur} but the problem remains open. In fact, as discussed in \cite{Balasubramanian:2011ur}, one of the main difficulties for putting forward a holographic proposal on momentum-space entanglement is the fact that this quantity is not well studied on the field theory side (specially for interacting theories). Hence, we hope our work can contribute to this issue.

\subsection*{Acknowledgements}

The authors are pleased to thank Veronika Hubeny, Mukund Rangamani, and Horatiu Nastase for discussions and comments on the draft. 
D.W.F.A. would like to acknowledge hospitality at the QMAP center of UC Davis, where part of this work was developed. 
D.W.F.A is supported  by the PDSE Program scholarship of CAPES – Brazilian Federal Agency for Support and Evaluation of Graduate Education within the Ministry of Education of Brazil, and by the CNPq grant 146086/2015-5.
G.C. acknowledges financial support from the Brazilian ministries MCTI and MEC. 

\bibliographystyle{JHEP}
\bibliography{refs}

\end{document}